\documentclass[twocolumn,10pt,journal]{IEEEtran}

\usepackage{cite}  %
\usepackage[tbtags]{amsmath}  %
\usepackage{amssymb}  %
\usepackage{graphicx}  %
\usepackage[caption=false]{subfig} %
\usepackage{soul}
\usepackage{array}
\usepackage{epstopdf} %
\usepackage{microtype} %
\usepackage{multirow} %
\usepackage{dcolumn} %
\usepackage{xfrac} %
\usepackage{mathrsfs} %

\usepackage{algorithm}
\usepackage[noend]{algpseudocode}

\usepackage{color}  %

\usepackage[framemethod=tikz]{mdframed}

\IEEEoverridecommandlockouts  %

\newcommand{\norm}[1]{{\left\lVert#1\right\rVert}}  %

\def\+{\texttt{+}}  %
\def\-{\texttt{-}}  %
\def\*{\mathtt{*}}  %

\def\kp{{k\texttt{+\!}}}  %
\def\km{{k\texttt{-\!}}}  %
\def\kpm{{k\scalebox{.6}{$\pm$}}}  %
\def\ko{{k\mathtt{*\!}}}  %

\def\kth{k^\text{th}}
\def\wok{{\!\hspace{.05em}\setminus{\!\hspace{.1em}k}}}  %

\def\plusone{\texttt{+}1}  %
\def\minusone{\texttt{-}1}  %
\def\pmone{\scalebox{.85}{$\pm$}1}

\def\eqpone{\,\texttt{=}\,\plusone}  %
\def\eqmone{\,\texttt{=}\,\minusone}  %
\def\eqpmone{\,\texttt{=}\,\pmone}  %

\begin{document}

\title{\fontsize{22.5}{24}\selectfont Achieving Near MAP Performance with an Excited Markov Chain Monte Carlo MIMO Detector}  %

\author{\normalsize
Jonathan C. Hedstrom, Chung Him (George) Yuen, Rong-Rong Chen, Behrouz Farhang-Boroujeny, \\
ECE department, University of Utah, USA \\
Email: \{jonathan.hedstrom, yuen, rchen, farhang\} @ece.utah.edu

\thanks{This work is supported in part by the National Science Foundation under grants 1449033 and 1632569.}
}
\maketitle

\begin{abstract}

We introduce a revised derivation of the bitwise Markov Chain Monte Carlo (MCMC) multiple-input multiple-output (MIMO) detector. The new approach resolves the previously reported high SNR stalling problem of MCMC without the need for hybridization with another detector method or adding heuristic temperature scaling factors. Another common problem with MCMC algorithms is the unknown convergence time making predictable fixed-length implementations problematic. When an insufficient number of iterations is used on a slowly converging example, the output LLRs can be unstable and overconfident. Therefore, we develop a method to identify rare slowly converging runs and mitigate their degrading effects on the soft-output information. This improves forward-error-correcting code performance and removes a symptomatic error floor in BER plots. Next, pseudo-convergence is identified with a novel way to visualize the internal behavior of the Gibbs sampler. An effective and efficient pseudo-convergence detection and escape strategy is suggested. Finally, the new excited MCMC (X-MCMC) detector is shown to have near maximum-a-posteriori (MAP) performance even with challenging, realistic, highly-correlated channels at the maximum MIMO sizes and modulation rates supported by the 802.11ac WiFi specification, 8$\times$8 MIMO 256 QAM.

\end{abstract}

\section{Introduction}

The use of spatial-multiplexing multiple-input multiple-output (MIMO) is increasingly being adopted in wireless protocols as higher spectral efficiency is needed to meet the capacity requirements of modern wireless networks \cite{halperin2010802,boccardi2012multiple}. It has the potential to linearly increase spectral reuse and capacity as the number of streams increases. Compared to massive-MIMO, more moderately sized spatial-multiplexing MIMO systems have many potential benefits including higher single-device throughput in scattering environments, small scale system capabilities appropriate for home use and small cells, and immediate availability in several common wireless standards. Up to 8-stream MIMO is already defined in the 802.11ac WiFi and LTE Advanced Release 10 protocols. With these large sizes, the performance and complexity scaling of the MIMO detector, which separates the mixed together transmitted streams at the receiver, is extremely important. The maximum-a-posteriori (MAP) detector has an exponentially increasing complexity which makes it unimplementable for all but the most trivial of cases. Zero forcing (ZF) and minimum mean square error (MMSE) detectors have low complexity but suffer from noise enhancement which degrades their performance sufficiently to negate the potential spectral efficiency benefits of MIMO.

The sphere-decoding (SD) class of MIMO detectors are well known to have near MAP performance \cite{Hochwald2003}. Specifically the K-Best variations of sphere-decoding have been demonstrated in effective VLSI designs but their complexity increases quickly with the number of antennas, the number of transmitted bits per channel use, and the list size $\kappa$ \cite{Guo2006}. Therefore, the search for alternative, lower complexity methods is ongoing.

Markov Chain Monte Carlo (MCMC) has been shown to have near optimal performance at low signal-to-noise-ratio (SNR) and to have efficient hardware implementations \cite{Farhang2006,Laraway2009}. It uses a random walk through the permutations of the transmitted bit sequence to estimate the posterior probability distribution and accordingly generate soft-output information. Unlike other detectors, the MCMC detector thus far has had the undesirable behavior that its performance degrades at higher values of SNR \cite{mao2007markov}. This issue is caused by stalling at high SNR. Most of the attempted solutions to this problem can be grouped into either hybridization or temperature scaling approaches. The hybridization schemes combine a method with good, low complexity performance at high SNR with MCMC to combine the best traits of both. Examples are ZF, MMSE, and sphere-decoding initialized MCMC methods \cite{peng2008low,yuan2011hardware}. The temperature scaling approaches are so called because they recognize that the probabilities generated by the Gibbs sampler are too cold, resulting in slow convergence. A linear scaling coefficient is therefore applied to artificially increase the noise temperature. This coefficient must be heuristically optimized depending on system parameters and SNR \cite{hansen2009near,senst2011rao,auras2014vlsi}, though a recent derivation of a near optimal value is showing promise under some testing conditions \cite{hassibi2014optimized}.

There are three main contributions presented in this paper. First, we present an improved derivation of the MCMC detector's Gibbs sampler which resolves the high SNR stalling problems that have been previously reported. Our new derivation uses a new and more accurate system model when a Gibbs sampler is far from the correct solution, accounting for both noise and bits-in-error. This results in what is effectively a dynamic temperature scaling coefficient. Then, using the ideas developed in this derivation, the problem of the unpredictable convergence time of the MCMC detector is explored. This leads to development of soft-output conditioning based on the quality of the list of samples generated by the random walk. This allows the detection of poor convergence and optionally extending the search with more iterations or terminating with a fixed number of iterations and decreasing the output confidence. Moderating the overconfident outputs is found to improve the performance of the forward-error-correcting (FEC) decoder. We also note that since the output conditioning is derived using the sample list, this method is applicable to other list based algorithms such as list sphere-decoding and K-Best.

The third contribution of this paper is identification and remediation of pseudo-convergence of the Gibbs sampler \cite{brooks2011handbook}. A novel way to visualize the Gibbs sampler behavior is used which provides insight into both  pseudo-convergence and stalling. To mitigate the resulting correlated sampling, we propose a simple motion based detection strategy and a 1-bit forced change procedure which sufficiently excites the Gibbs sampler to allow new uncorrelated samples to be made.

The excited MCMC (X-MCMC) detector that is proposed in this paper is comprised of the excited Gibbs sampler, output LLR conditioning, and pseudo-convergence remediation through additional Gibbs excitation. The new X-MCMC detector achieves near Max-MAP performance even at the largest 802.11ac WiFi rates with 8$\times$8 MIMO and 256 QAM modulation. This is demonstrated on the challenging, highly correlated WiFi TGn Model-D channel \cite{ieee2004tgn}, which is much more difficult to attain near MAP performance on than with uncorrelated i.i.d.\ Gaussian channels \cite{hedstrom2017icc}.
This paper is organized as follows. In Section~\ref{sec:system_model} we present the system model and some specialized notation used throughout the paper. A review of a typical MCMC derivation along with background information is provided in Section~\ref{sec:mcmc}. The first two contributions of this paper are explained in Section~\ref{sec:xmcmc}; in Subsection~\ref{sec:xmcmc - gibbs excitation} the excited Gibbs sampler is presented and our proposed method of output LLR conditioning is discussed in Subsection \ref{sec:xmcmc - llr output conditioning}. Our third contribution, the pseudo-convergence of Gibbs samplers and remedy to it, is discussed in Section~\ref{sec:pseudo-convergence}. Simulation results and a complexity analysis are presented in Section~\ref{sec:results}, and the conclusions of the paper are made in Section~\ref{sec:conclusion}.
\section{System Model}\label{sec:system_model}

The goal of a spatial-multiplexing MIMO transceiver is to exploit the spatial multipath of the environment to support overlapping data streams. This concept allows reuse of the spectrum and therefore has the potential to dramatically increase the data rates and capacity of wireless networks.

To perform data transmission with spatial-multiplexing, the transmitter sends independent data streams simultaneously on a set of $N_\text{t}$ transmit antennas which are then received at a set of $N_\text{r}$ receive antennas. In an orthogonal frequency division multiplexing (OFDM) system, for each subcarrier, this can be represented with a flat-fading frequency-domain system model as
\begin{equation}\label{eqn:system_model}
\mathbf{y} = \mathbf{H}\mathbf{s} + \mathbf{n}
.
\end{equation}
Here, $\mathbf{y}$ is the received signal vector, $\mathbf{H}$ is a slow flat fading complex channel matrix containing the pairwise gain and phase between antennas, $\mathbf{s}$ is the vector of transmitted complex constellation symbols mapped from the bit sequence $\mathbf{x}$, and $\mathbf{n}$ is a noise vector. Assuming the transmit and receive side have the same number of antennas $N_t=N_r=N$, the dimensions of the corresponding vectors and matrices are $N\!\times\!1$ and $N\!\times\!N$, respectively. Note that often it is easier in hardware designs to use a real-valued system model which this papaer is fully compatible with, see \cite{Laraway2009}. The channels for this paper are produced with the method described by the WiFi TGn Model-D specification \cite{ieee2004tgn} which creates a correlated channel matrix $\mathbf{H}$. The TGn-D channel is much more challenging for the detector compared to the independent and identically distributed (i.i.d.) complex Gaussian channel typically used in the literature \cite{Hochwald2003,Guo2006,Farhang2006,Laraway2009,hassibi2014optimized}. The bit vector $\mathbf{x}$ may be described as comprising of 1's and 0's or equivalently $\plusone$'s and $\minusone$'s, depending on context. In principle, $\mathbf{s}$ can be a vector of any complex modulated symbols, but here we use the quadrature amplitude modulated (QAM) symbols defined in the IEEE 802.11ac specification. The elements of $\mathbf{n}$ are assumed to be i.i.d.\ complex Gaussian random variables with variance of $\sigma_n^2$ per each real and imaginary dimension.

The primary challenge in a MIMO receiver, and the focus of this paper, is to accurately and efficiently estimate the $K=N\log_2(N_\text{qam})$ simultaneously transmitted bits in $\mathbf{x}$ per realization, where $N_\text{qam}$ is the QAM constellation size.

\begin{figure}[t]
    \centering
    \vspace{-2mm}
    \includegraphics{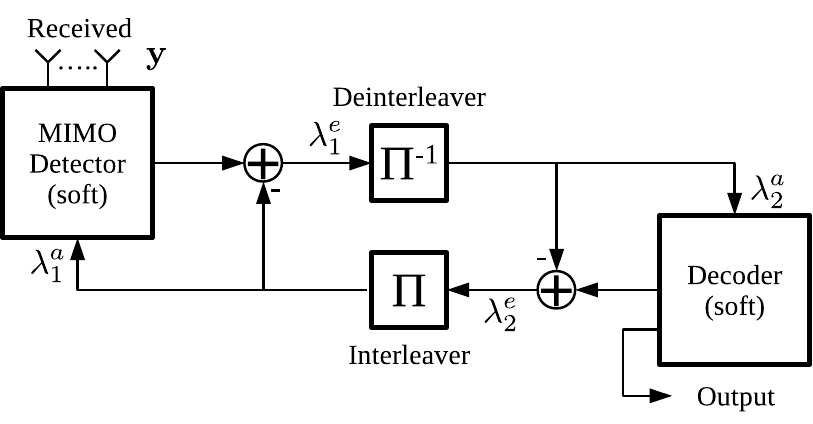}
    \vspace{-2mm}
    \caption{Turbo loop structure with a MAP, MCMC, X-MCMC, or K-Best MIMO detector iteratively exchanging soft information with a forward error correction decoder.}
    \label{fig:turbo_loop}
    \vspace{-.5\baselineskip}  %
\end{figure}

The MIMO detectors discussed in this paper are all compatible with turbo iterations, the exchange of soft-information between the detector and decoder as in Fig.~\ref{fig:turbo_loop}. This allows for the iterative joint detection of the signal for enhanced performance. The soft-information is in the form of log likelihood ratios (LLRs) and represented by $\lambda^a$ and $\lambda^e$ for their \textit{a priori} input and extrinsic output versions.

\subsection*{Notation}

In the equations that follow, some specialized notation is used for compactness and clarity. Vectors and matrices are expressed with bold fonts and the latter are capitalized. The removal of the $\kth$ element of a vector is shown with set notation as $\{\boldsymbol{\cdot}\}^\wok$. A variable or vector derived from the bit sequence $\mathbf{x}$ with the $\kth$ bit forced to a one or zero is shown with $\{\cdot\}^\kp$ and $\{\cdot\}^\km$, respectively. When two nearly identical equations are needed differing only in use of $\kp\,$ or $\km\,$, $k\scalebox{.8}{$\pm$}$ is used to represent both versions. If the $\kth$ bit is forced to the correct transmitted value, either one or zero, it is shown with $\{\cdot\}^\ko$.
\section{MCMC Detector}\label{sec:mcmc}

The MCMC MIMO detector estimates the output log likelihood ratio (LLR) by means of Monte Carlo sampling \cite{Farhang2006}. This can be thought of as a method to identify a list of important bit permutation samples to approximate the full permutation list in the MAP detector. The challenge is to make the process computationally efficient by using an easy to calculate short list that accurately captures the statistics of the full permutation list. We select the bitwise method described in \cite{Laraway2009} as our foundational algorithm because it has been shown to be efficiently implementable in hardware.

The two main components of the MCMC detector is the Gibbs sampler and the LLR output calculation. The Gibbs sampler starts with an initial estimate of the transmitted bit sequence, either randomly selected or initialized with prior information. It then cycles through the bits, calculating the probability of a bit being a one or zero and uses it to weight a random decision to change the bit. Each cycle through the bits is an iteration. The list of permutations visited by the Gibbs sampler is used to calculate the output LLR.

To make the differences between our improved excited MCMC detector and the original bitwise MCMC detector clear, here, we will show the main steps in the original MCMC's Gibbs sampler and output LLR derivation. After a brief summary of the algorithm, we will provide some detail of the stalling problem and the MMSE initialized MCMC detector variation used in comparisons of Section \ref{sec:results}.

\subsection{Gibbs Sampler}\label{sec:mcmc - gibbs derivation}

The Gibbs sampler is at the core of the MCMC algorithm. It is used in difficult multi-variate posterior probability estimation problems when sampling probabilities jointly across all variables is too complex. It cycles across the variables, calculating probabilities conditioned on all other variables being fixed to the current state. In the bitwise MCMC MIMO detector, this means that the Gibbs sampler calculates the probability of a specific bit $x_k$ being a $\plusone$ or $\minusone$ conditioned on the current state of $\mathbf{x}^\wok$. Therefore, the probability $P(x_k\eqpone|\mathbf{y},\mathbf{x}^\wok,\boldsymbol{\lambda}^a)$ is needed which we refer to henceforth as $P_\text{gibbs}$ for brevity.

To derive $P_\text{gibbs}$ we begin with the definition of the LLR for the $\kth$ bit at the current Gibbs sampler state $\mathbf{x}$
\begin{equation}\label{eqn:mcmc gamma definition}
\gamma_k
= \ln \frac{P(x_k\eqpone|\mathbf{y},\mathbf{x}^\wok,\boldsymbol{\lambda}^a)}{P(x_k\eqmone|\mathbf{y},\mathbf{x}^\wok,\boldsymbol{\lambda}^a)}.
\end{equation}
Then, by noting that
\begin{equation}
P(x_k\eqmone|\mathbf{y},\mathbf{x}^\wok,\boldsymbol{\lambda}^a) = 1 - P(x_k\eqpone|\mathbf{y},\mathbf{x}^\wok,\boldsymbol{\lambda}^a)
,
\end{equation}
we can rearrange to have the definition of the Gibbs probability as
\begin{equation} \label{eqn:prob gibbs}
P_\text{gibbs} = P(x_k\eqpone|\mathbf{y},\mathbf{x}^\wok,\boldsymbol{\lambda}^a) = \frac{1}{1+e^{-\gamma_k}}.
\end{equation}

Next, Bayes' Theorem is applied to $\gamma_k$ to separate the contribution of the prior $\boldsymbol{\lambda}^a$. The result is
\begin{equation} \label{eqn:mcmc gamma bayes}
\begin{split}
\gamma_k
&= \ln \frac
    {p(\mathbf{y} | x_k\eqpone,\mathbf{x}^\wok,\boldsymbol{\lambda}^a) P(x_k\eqpone|\mathbf{x}^\wok,\boldsymbol{\lambda}^a) p(\mathbf{y})}
    {p(\mathbf{y} | x_k\eqmone,\mathbf{x}^\wok,\boldsymbol{\lambda}^a) P(x_k\eqmone|\mathbf{x}^\wok,\boldsymbol{\lambda}^a) p(\mathbf{y})}\\
&= \ln \frac
    {p(\mathbf{y} | x_k\eqpone,\mathbf{x}^\wok) P(x_k\eqpone | \lambda^a_k) }
    {p(\mathbf{y} | x_k\eqmone,\mathbf{x}^\wok) P(x_k\eqmone | \lambda^a_k) }\\
&= \ln \frac 
    {p(\mathbf{y}|x_k\eqpone,\mathbf{x}^\wok)}
    {p(\mathbf{y}|x_k\eqmone,\mathbf{x}^\wok)} +\lambda_k^a
\end{split}
\end{equation}
where in the second line $\boldsymbol{\lambda}^a$, $\mathbf{x}^\wok$, and $\boldsymbol{\lambda}^{a,\wok}$ are removed from their respective conditionals because of the independence among the $\mathbf{x}$ bits created by the interleaving effect in the turbo loop. The final line follows from the definition of the {\em a priori} LLR $\boldsymbol{\lambda}^a$.

The system model is now used to provide the probability of the received sequence $\mathbf{y}$ given $\mathbf{x}$ and noise with variance $\sigma_n^2$ per dimension. This leads to
\begin{equation}\label{eqn:mcmc p(y|x)}
p(\mathbf{y}|\mathbf{x}) = \left(2\pi\sigma_n^2\right)^{-N} \exp\left(-\frac{\norm{\mathbf{y}-\mathbf{H}\mathbf{s}}^2}{2\sigma_n^2}\right)
.
\end{equation}
Substituting (\ref{eqn:mcmc p(y|x)}) into (\ref{eqn:mcmc gamma bayes}) and simplifying the result, we get
\begin{equation}\label{eqn:mcmc gamma gaussian}
\gamma_k = \frac{1}{2\sigma_n^2} \left(\norm{\mathbf{y}-\mathbf{H}\mathbf{s}^\km}^2 - \norm{\mathbf{y}-\mathbf{H}\mathbf{s}^\kp}^2 \right) + \lambda^{a}_k
.
\end{equation}
This may be used with (\ref{eqn:prob gibbs}) to calculate the needed $P_\text{gibbs}$ probability for the Gibbs sampler. Note that this will only be an accurate calculation if AWGN noise with variance $\sigma_n^2$ is the only contributer to the error residual in $\mathbf{y}-\mathbf{H}\mathbf{s}$. This assumption is revisited in the derivation of the X-MCMC detector in Section~\ref{sec:xmcmc} where we find that error in the Gibbs sampler's current state of $\mathbf{s}$ introduces additional error that should be accounted for.

\subsection{Output LLR}\label{sec:mcmc - output derivation}

To calculate the extrinsic output LLR $\boldsymbol{\lambda}^e$ we first derive the MAP output LLR. The definition of LLR for the $\kth$ bit is the log ratio of probabilities of $x_k$ being a $\plusone$ or $\minusone$. This is expressed as
\begin{equation}\label{eqn:mcmc llr definition}
\begin{split}
\lambda_{k,\text{MAP}}^e
&= \ln \frac{P( x_k \eqpone|\mathbf{y},\boldsymbol{\lambda}^{a,\wok} )}{P( x_k \eqmone|\mathbf{y},\boldsymbol{\lambda}^{a,\wok} )} \\
&= \ln \frac
{\displaystyle\sum_{\mathbf{x}\in\mathbb{X}^\kp} \left( p(\mathbf{y}|\mathbf{x})\prod_{j \ne k} P(x_j|\lambda_j^a) \right)}
{\displaystyle\sum_{\mathbf{x}\in\mathbb{X}^\km} \left( p(\mathbf{y}|\mathbf{x})\prod_{j \ne k} P(x_j|\lambda_j^a) \right)}
\end{split}
\end{equation}
where $\mathbb{X}^\kp$ and $\mathbb{X}^\km$ are the sets of all permutations of $\mathbf{x}$ with the $\kth$ bit forced to a $\plusone$ and $\minusone$, respectively. To simplify (\ref{eqn:mcmc llr definition}), we recall from \cite{hagenauer1996iterative} that
\begin{equation}
P(x_j|\lambda_j^a) = \frac{e^{-\lambda_j^a/2}}{1+e^{-\lambda_j^a}} e^{x_j\lambda_j^a/2} = A(\lambda_j^a) e^{x_j\lambda_j^a/2}
.
\end{equation}
Since the coefficient $A(\lambda_j^a)$ is independent of $x_j$, it may be separated from the summation and canceled. This leads to
\begin{equation}\label{eqn:mcmc llr A removed}
\begin{split}
\lambda_{k,\text{MAP}}^e &= \ln \frac
{\displaystyle\sum_{\mathbf{x}\in\mathbb{X}^\kp} \left( p(\mathbf{y}|\mathbf{x}) \exp \sum_{j \ne k} \frac{1}{2}x_j\lambda_j^a \right)}
{\displaystyle\sum_{\mathbf{x}\in\mathbb{X}^\km} \left( p(\mathbf{y}|\mathbf{x}) \exp \sum_{j \ne k} \frac{1}{2}x_j\lambda_j^a \right)}
.
\end{split}
\end{equation}
Finally, the max-log approximation to the Jacobian logarithm \cite{robertson1995comparison} is used to simplify (\ref{eqn:mcmc llr A removed}) to 
\begin{equation}\label{eqn:llr output maxlog map}
\begin{split}
\lambda_{k,\text{Max-MAP}}^e \approx
 &\frac{1}{2} \max_{\mathbf{x}\in\mathbb{X}^\kp} \!\! \left( -\frac{1}{\sigma_n^2}\norm{\mathbf{y}-\mathbf{H}\mathbf{s}^\kp}^2 + \mathbf{x}^\wok \!\cdot\! \boldsymbol{\lambda}^{a,\wok} \right) \\
-&\frac{1}{2} \max_{\mathbf{x}\in\mathbb{X}^\km} \!\! \left( -\frac{1}{\sigma_n^2}\norm{\mathbf{y}-\mathbf{H}\mathbf{s}^\km}^2 + \mathbf{x}^\wok \!\cdot\! \boldsymbol{\lambda}^{a,\wok} \right)
\end{split}
\end{equation}
where $\mathbf{x}^\wok \!\cdot\! \boldsymbol{\lambda}^{a,\wok}$ is a vector dot product. This approximation is commonly used by approximate-MAP MIMO detectors as it results in a minor loss of performance while significantly decreasing algorithm complexity \cite{Hochwald2003}.

To calculate the output LLR with MCMC, the max-log MAP calculation of (\ref{eqn:llr output maxlog map}) is used but with the set $\mathbb{X}$ replaced with the list $\text{Z}$ of the sampled permutations. This, with some minor rearrangements leads to
\begin{equation}\label{eqn:llr output mcmc}
\begin{split}
\lambda_{k,\text{MCMC}}^e \approx
 &\frac{1}{2}\min_{\mathbf{x}\in\text{Z}^\km} \!\! \left( \frac{1}{\sigma_n^2}\norm{\mathbf{y}-\mathbf{H}\mathbf{s}^\km}^2 - \mathbf{x}^\wok \!\cdot\! \boldsymbol{\lambda}^{a,\wok} \right) \\
-&\frac{1}{2}\min_{\mathbf{x}\in\text{Z}^\kp} \!\! \left( \frac{1}{\sigma_n^2}\norm{\mathbf{y}-\mathbf{H}\mathbf{s}^\kp}^2 - \mathbf{x}^\wok \!\cdot\! \boldsymbol{\lambda}^{a,\wok} \right)
\end{split}
\end{equation}
where the $\max()$ operation has been changed to an equivalent $\min()$ operation to put the equation in a more intuitive cost function minimization form.

\subsection{Summary}

The MCMC MIMO detector algorithm starts with an initial sequence of bits $\mathbf{x}$. It then cycles across the $K$ bits in a bitwise fashion for $N_\text{iter}$ iterations, using the probabilities calculated with (\ref{eqn:prob gibbs}) and (\ref{eqn:mcmc gamma gaussian}) to determine state transitions. This process effectively guides the Gibbs sampler towards the more important regions. After a sufficient quantity of samples have been taken in this manner, the algorithm approximates the max-log MAP detector's extrinsic output LLR $\lambda^e_k$ in (\ref{eqn:llr output maxlog map}) with (\ref{eqn:llr output mcmc}). This procedure is outlined in Algorithm~\ref{alg:gibbs} for one Gibbs sampler, though $N_\text{gibbs}$ can be used in parallel to increase sampling speed and increase sample diversity.

\begin{algorithm}
\caption{Basic MCMC Gibbs Sampler}
\label{alg:gibbs}
\begin{algorithmic}[1]
\State Initialize $\mathbf{x}$.
\For{$1$ to $N_\text{iter}$}
    \For{$k=1$ to $N\log_2(N_\text{qam})$}
        \State Update $\min_{\mathbf{x}\in\text{Z}^\kp}(\;)$ and $\min_{\mathbf{x}\in\text{Z}^\km}(\;)$ for (\ref{eqn:llr output mcmc}).
        \State Calculate $P_\text{gibbs}$ with (\ref{eqn:prob gibbs}) and (\ref{eqn:mcmc gamma gaussian}).
        \State Generate a uniform random variable $0\le r \le 1$.
        \If{$r < P_\text{gibbs}$}
            \State $x_k \leftarrow \plusone$
        \Else
            \State $x_k \leftarrow \minusone$
        \EndIf
    \EndFor
\EndFor
\State Compute output LLR with (\ref{eqn:llr output mcmc}).
\end{algorithmic}
\end{algorithm}

\subsection{Stalling Problem}\label{sec:mcmc - stalling}

At high SNR, the MCMC detector derived here is known to stall. This means that the rate at which new samples are generated per iteration approaches zero \cite{zhu2005mimo}. This results in slow convergence and an enormous number of iterations needed to reach near MAP performance, potentially more than the calculations needed for MAP itself. To help understand stalling more thoroughly and the improvements proposed in the subsequent sections, a detailed view of the underlying Gibbs samplers is presented that will lead to some insight.

\begin{figure}[t]
    \centering
    \vspace{-2mm}
    \subfloat[Correct behavior at low SNR ($E_b/N_0=6\text{dB}$).\label{fig:16qam low snr}]{\includegraphics[trim=0mm 1mm 0mm 0mm,clip,width=2.5in]{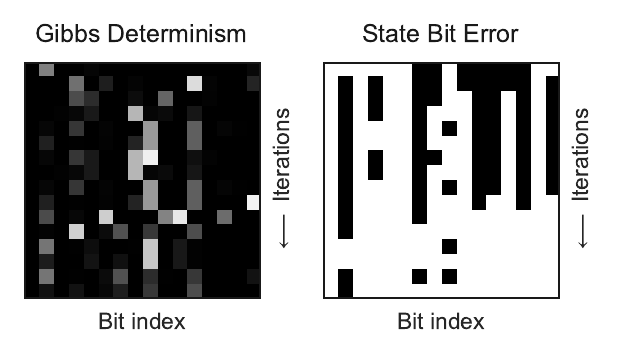}}
    \vfill
    \subfloat[Stalling at high SNR ($E_b/N_0=12\text{dB}$).\label{fig:16qam high snr}]{\includegraphics[trim=0mm 1mm 0mm 0mm,clip,width=2.5in]{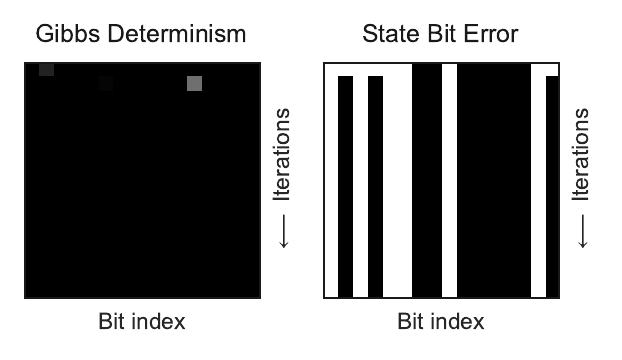}}
    \caption{Gibbs sampler detail with the original randomly initialized MCMC algorithm. Both subfigures have the exact same bit sequence, channel, seeded random number generator, and additive noise. The only difference is in noise magnitude which instigates the high SNR stalling problem. Parameters: 4 antennas, 16 QAM, $N_\text{iter} = 16$, WiFi TGn Model-D channel.\label{fig:16qam}}
\end{figure}

In Fig.~\ref{fig:16qam}, we see the internal behavior of a pair of randomly initialized Gibbs samplers working on a 4$\times$4 MIMO system with a 16 QAM symbol constellation, thus $4\log_2(16)=16$ bits per $\mathbf{x}$. The only difference between the subfigures is the magnitude of AWGN noise, 6 dB vs 12 dB $E_b/N_0$, which is sufficient to strongly instigate the high SNR stalling problem. Each subplot has the bit index $k$ over the horizontal axis from left to right, and the Gibbs iterations starting at the top and descending over time. Accordingly, the Gibbs sampler moves from left to right and top to bottom.

The left subplots show the probability of deterministic and non-deterministic flips where
\begin{equation}\label{eq:determinism}
\text{determinism} = |2P_\text{gibbs}-1| .
\end{equation}
With a gray-scale color mapping the determinism is shown as $\text{black}=1$ (fully deterministic) and $\text{white}=0$ (fully random).

The right subplots show the error in the Gibbs state $\mathbf{x}$ when compared to the true transmitted bit sequence, where black indicates a bit error. State error is used instead of simply $\mathbf{x}$ because it simultaneously shows if the state is changing and where the random walk is relative to the transmitted bit sequence. 

In Fig.~\ref{fig:16qam}, we see the curious trend that at low SNR the Gibbs sampler probability is moderate and the sampler quickly converges to having only 1-bit error, whereas at high SNR the Gibbs probabilities are persistently at extreme values and no change is seen after the first iteration. 

Stalling is not seen under all conditions. If a smaller QAM size of 2 to 16 is used at a low SNR, near the BER cliff, stalling will not be observed \cite{kumar2011near,shi2004markov}. Another condition where the full extent of stalling may not be obviously observed is when using well-conditioned, Rayleigh channels with uncorrelated i.i.d.\ Gaussian channel matrix elements \cite{hedstrom2017icc}. Real-world channels, with correlation between antennas, display stalling much more strongly than the uncorrelated channels often used in the literature to simulate MCMC performance \cite{Guo2006,Laraway2009,mao2007markov,hassibi2014optimized}. There are many variations of the MCMC detector which attempt to mitigate the high SNR stalling problem. Most can be broadly classified as either hybrid or temperature scaling methods.

The idea behind the hybrid approaches is to combine the MCMC detector, which has decreasing performance as SNR increases, with a complementary detector, which has low performance at low SNR but high performance at high SNR. There are many examples of this using the sphere-decoder such as \cite{peng2008low,yuan2011hardware} and with MMSE \cite{mao2007markov}. An interesting variation on this theme is to use successive-over-relaxation within each Gibbs sampler step calculation \cite{choi2016mcmc}. We note that although these methods can improve the performance compared to a standard MCMC sampler by itself, this class of detectors do not solve the high SNR stalling problem of the MCMC algorithm they use.

The temperature scaling class of MCMC detectors recognize that the superficial cause of stalling is that the $P_\text{gibbs}$ values calculated are too extreme. They suggest that, as the SNR increases, $\sigma_n^2$ acts as a gain term in (\ref{eqn:mcmc gamma gaussian}), therefore they heuristically add an additional scaling term $\alpha$ which either linearly scales or replaces $\sigma_n^2$ to counteract stalling \cite{hansen2009near,hassibi2014optimized,datta2012novel}. A variation on this idea includes methods which use manually selected parameters to adjust the inclusion of additional randomization into the probability calculation or bit selection \cite{kumar2011near,he2015low}.

Stalling will be further explained in the framework of the X-MCMC detector variation of Section~\ref{sec:xmcmc - gibbs excitation}.

\subsection{MMSE Initialized MCMC}\label{sec:mcmc - mmse}

To better understand the improvements made by the X-MCMC detector that will be introduced in Section~\ref{sec:xmcmc}, an MCMC method is needed for comparison. We have selected the MMSE initialized MCMC detector described in \cite{mao2007markov}. This is a hybrid type of detector since it combines MMSE, which has good performance at high SNR, with MCMC to compensate for the high SNR stalling effect. The reason why we have selected this as apposed to some of the other excellent QRD-hybrid or temperature scaling varieties in the literature is that those methods are generally loosely defined with parameters that need to be heuristically tuned to the application. The MMSE initialized variety is useful as a benchmark because it has an explicit implementation regardless of channel, SNR, number of antennas, or modulation order.

The only extension needed for the MMSE initialized MCMC hybrid algorithm is to initialize one of the parallel Gibbs samplers with the hard decision from an MMSE solution as in
\begin{equation}\label{eqn:mmse detector}
\hat{\mathbf{s}}_\text{MMSE} = \left( \mathbf{H}^\dagger\mathbf{H}+2\sigma_n^2\mathbf{I} \right)^{-1} \mathbf{H}^\dagger \mathbf{y}
\end{equation}
where $\dagger$ is the conjugate transpose. This method solves the issue where the MCMC detector does not converge for small QAM sizes at high SNR. It works because the MMSE solution will result in a correctly signed LLR output without any need of additional MCMC Gibbs iterations. For turbo iterations to work properly and to generate reasonable extrinsic information transfer (EXIT) charts, every iteration has one parallel Gibbs sampler initialized with the original MMSE solution.

As described in Section~\ref{sec:mcmc - stalling}, the MMSE initialized MCMC detector does not prevent stalling because it is a hybrid method, thus it is far from MAP performance when higher QAM sizes are used. Examples that show such behavior are presented in Section~\ref{sec:results}.

\section{X-MCMC Detector}\label{sec:xmcmc}

In this section, we present the two main components of the excited MCMC (X-MCMC) detector: the excited Gibbs sampler and LLR output conditioning. Both rely on the realization that there are error contributions from both noise and bits-in-error. Previous derivations have only included AWGN noise in their derivations. We found that the inclusion of error in the statistical model completely solves the high SNR stalling problem covered in Section \ref{sec:mcmc - stalling}. It also provides a method to detect and mitigate poorly converged and overconfident output LLRs that otherwise confuse the decoder, potentially creating error floors in BER plots.

Previously in \cite{hedstrom2017icc}, we used a heuristic explanation to justify the dynamic scaling used in the X-MCMC detector. Simulations were matched to 8$\times$8 MIMO testbed measurements with near max-log MAP performance. Those ideas will be expanded on here with a thorough theoretical understanding and some algorithmic improvements.

In the discussions and derivations that follow, the concept of distance is repeatedly used. It is the closeness of the current state $\mathbf{x}$ to the transmitted sequence. To simplify its use, it will be defined as the square Euclidean distance
\begin{equation}\label{eqn:d}
d = \norm{ \mathbf{y} - \mathbf{H}\mathbf{s} }^2
\end{equation}
where $\mathbf{s}$ is the complex symbol mapped version of the bit state $\mathbf{x}$.
If the $\kth$ bit of $\mathbf{x}$ is forced to a one or zero, then this can be indicated on all dependent variables and vectors with a superscript $\kp\,$ or $\km\,$ as in
\begin{equation}\label{eqn:dkpm}
d^\kpm = \norm{ \mathbf{y} - \mathbf{H}\mathbf{s}^\kpm }^2
.
\end{equation}

\subsection{Excited Gibbs Sampler}\label{sec:xmcmc - gibbs excitation}

The cause of Gibbs sampler stalling in the MCMC detector is constant production of extreme $P_\text{gibbs}$ values close to 0\% or 100\%. This creates a nearly deterministic walk which quickly stalls. Extreme probabilities are readily caused by (\ref{eqn:prob gibbs}) and (\ref{eqn:mcmc gamma gaussian}). At just $|\gamma_k|=3$ there is a mere 2\% chance of a non-deterministic transition. In simulations of 4-antenna 64 QAM systems, $\gamma_k$ values greater than 10 are commonly seen and values greater than 100 are not unusual at high SNRs.

In the MCMC derivation of Section~\ref{sec:mcmc}, an assumption is implicitly made that the only form of error is from the channel model's AWGN noise. It is a useful assumption because it allows the substitution of a Gaussian probability distribution into (\ref{eqn:mcmc gamma bayes}) that simplifies to the explicit $\gamma_k$ of (\ref{eqn:mcmc gamma gaussian}). However, this causes problems since the incorrect bits in $\mathbf{x}$ also contribute to error and in general are much larger contributors, often by a factor of 1000 at high SNRs. This provides insight as to why the stalling problem only appears at high SNR, since at low SNRs the condition is less strongly violated. Therefore, the goal here is to find a way to calculate the needed probabilities in $\gamma_k$ without this assumption.

We begin the derivation of the excited Gibbs sampler with (\ref{eqn:mcmc gamma bayes}) from the original MCMC derivation.
\begin{equation} \label{eqn:xmcmc gamma bayes}
\begin{split}
\gamma_k
&= \ln \frac 
    {p(\mathbf{y}|x_k\eqpone,\mathbf{x}^\wok)}
    {p(\mathbf{y}|x_k\eqmone,\mathbf{x}^\wok)} +\lambda_k^a
\end{split}
\end{equation}
The implicit statistical model used for the probability distributions in the original MCMC derivation is
\begin{equation} \label{eqn:statistical model}
\mathbf{n} = \mathbf{y}-\text{H}\mathbf{s}
.
\end{equation}
But a more accurate model is needed when the Gibbs sampler is far from the true transmitted bit sequence. The error $\mathbf{e}^\ko$ must include both AWGN interference and the incorrect bits in the state. This is captured with the new model
\begin{equation} \label{eqn:error model}
\mathbf{e}^\ko = \mathbf{y}-\text{H}\mathbf{s}^\ko = \mathbf{H} \left( \mathbf{s}_\text{tx} -\mathbf{s}^\ko \right) + \mathbf{n}
\end{equation}
where $\mathbf{s}_\text{tx}$ is the transmitted symbol vector and $\mathbf{s}^\ko\,$ is the current Gibbs state with the $\kth$ bit of $\mathbf{x}$ correct, i.e.\ matches the respective transmitted bit. This model does not include error from the $\kth$ bit in $\mathbf{s}$ so that the distributions of the numerator and denominator of (\ref{eqn:xmcmc gamma bayes}) will be the same.

It is not generally possible to separate the contributions of $x_k$ and $\mathbf{x}^\wok$ to the error, so in the steps that follow we will initially assume that an oracle has access to such an error metric which excludes the $\kth$ bit and finish with approximations sufficient for development of a detector that works well in practice.

To use the new error model, we first note that a Gaussian distribution is an appropriate approximation for the elements of $\mathbf{e}^\ko$. Not considering the contribution of the Gaussian noise, each element is generated from the dot product of a row of the known $\mathbf{H}$ and an unknown column of symbol errors. Therefore, each real and imaginary component of $\mathbf{e}^\ko$ is created from the sum of $4N$ uniform random variables, where the number of antennas has been multiplied by two for the complex nature of the symbols and two for the contribution of the independent $s_\text{tx}$ and $s^\ko$ symbols. For the smallest MIMO size, this results in an $n=8$ Irwin-Hall distribution which is approximately Gaussian and improves as the number of antennas increases. Though our QAM symbols take discrete values, the noise and known $\mathbf{H}$ randomly shift and scale the symbols sufficiently to retain the Gaussian approximation.

Here, we have assumed that $\mathbf{s}_\text{tx}$ and $\mathbf{s}^\ko$ are uncorrelated which is valid when far from convergence. As the Gibbs sampler converges and the symbols become more highly correlated, the true distribution becomes narrower than a Gaussian, thus, a Gaussian assumption results in underconfident probabilities near convergence. In Section~\ref{sec:pseudo-convergence}, we will find that pseudo-convergence is a stronger effect leading to overconfident probabilities during convergence. Therefore, the slight underconfidence here can be safely ignored.

Using a Gaussian distribution for the elements of $\mathbf{e}^\ko$ with variance $\sigma_\ko^2$ per dimension results in
\begin{equation}\label{eqn:xmcmc p(y|x)}
p(\mathbf{y}|x_k \eqpmone,\mathbf{x}^\wok) = \left(2\pi\sigma_\ko^2\right)^{-N} \exp\left(-\frac{\norm{\mathbf{y}-\mathbf{H}\mathbf{s}^\kpm}^2}{2\sigma_\ko^2}\right).
\end{equation}
Then, by substituting this Gaussian model into (\ref{eqn:xmcmc gamma bayes}), we have
\begin{equation}\label{eqn:gamma sigma_k*}
\gamma_k = \frac{d^\km - d^\kp}{2\sigma_\ko^2} + \lambda^{a}_k
.
\end{equation}

A single sample estimate of the error variance $\sigma_\ko^2$ can be made with the current error vector $\mathbf{e}^\ko$ as in
\begin{equation} \label{eqn:sigma_k* oracle}
\sigma_\ko^2 \approx \frac{\norm{\mathbf{e}^\ko}^2}{2N} = \frac{d^\ko}{2N}
\end{equation}
where we have noted that the squared norm of the error vector is equivalent to having the distance $d$ with the $\kth$ bit known to be correct, $d^\ko=\norm{\mathbf{y}-\mathbf{H}\mathbf{s}^\ko}^2$. We refer to this estimation of the error variance as the {\em oracle} method since it uses knowledge that is not generally available to the algorithm. Next, we will develop several potential approximations to the {\em oracle}.

We begin by using the confidence that either $d^\kp$ or $d^\km$ has the $\kth$ bit correct to generate a weighted average estimate $\hat{d}^\ko$.

\begin{equation} \label{eqn:dk* weighted derive}
\begin{split}
\hat{d}^\ko &= d^\kp P(x_k \eqpone|\mathbf{x}^\wok,\mathbf{y}) + d^\km P(x_k \eqmone|\mathbf{x}^\wok,\mathbf{y}) \\
&= \quad\frac{d^\kp p(\mathbf{y}|x_k \eqpone,\mathbf{x}^\wok)P(x_k \eqpone|\mathbf{x}^\wok)} {p(\mathbf{y}|\mathbf{x}^\wok)} \\
&\quad\,+\frac{d^\km p(\mathbf{y}|x_k \eqmone,\mathbf{x}^\wok)P(x_k \eqmone|\mathbf{x}^\wok)} {p(\mathbf{y}|\mathbf{x}^\wok)} \\
&= \frac{d^\kp p(\mathbf{y}|x_k \eqpone,\mathbf{x}^\wok) + d^\km p(\mathbf{y}|x_k \eqmone,\mathbf{x}^\wok)} {2p(\mathbf{y}|\mathbf{x}^\wok)}
\end{split}
\end{equation}
where Bayes' rule is first applied and then $P(x_k \eqpmone|\mathbf{x}^\wok)=1/2$ because of the independence between bits. This estimate can be bounded by the minimum and mean of $d^\kpm$ since $P(x_k \eqmone|\mathbf{x}^\wok,\mathbf{y})=1-P(x_k \eqpone|\mathbf{x}^\wok,\mathbf{y})$ and the larger probability will correspond to the smaller distance.
\begin{align}
\hat{d}^\ko_\text{min} = \min \left( d^\kp,d^\km \right) \le \hat{d}^\ko& \label{eqn:dp min}\\
\hat{d}^\ko& \le \frac{1}{2} \left( d^\kp+d^\km \right) = \hat{d}^\ko_\text{mean} \label{eqn:dp mean}
\end{align}
These bounds are useful rough approximations to $\hat{d}^\ko$ as they are readily available to the Gibbs sampler and are computationally efficient. We will refer to these as the {\em min} and {\em mean} approximations.

A more accurate approximation to the weighted average of (\ref{eqn:dk* weighted derive}) can be made by first replacing the denominator, which would require the unknown $\mathbf{x}^\ko$, with the average of the probabilities with $x_k\eqpmone$.
\begin{equation} \label{eqn:p(y|x*) estimate}
p(\mathbf{y}|\mathbf{x}^\wok) = \frac{1}{2} \left( p(\mathbf{y}|x_k \eqpone,\mathbf{x}^\wok) + p(\mathbf{y}|x_k \eqmone,\mathbf{x}^\wok) \right)
\end{equation}
Then, single sample estimates of the variances $\sigma_\kpm^2\approx d^\kpm/2N$ are used to approximate the Gaussian distributions
\begin{equation}\label{eqn:weighted gaussian}
\begin{split}
p(\mathbf{y}|x_k \eqpmone,\mathbf{x}^\wok)
&\approx \left(2\pi\sigma_\kpm^2\right)^{-N} \exp \left( -\frac{d^\kpm}{2\sigma_\kpm^2} \right) \\
&\approx \left( \pi\frac{d^\kpm}{N} \right)^{-N} \! \exp \left( -\frac{d^\kpm}{d^\kpm/N} \right)
.
\end{split}
\end{equation}
This estimate is too crude to be used directly in (\ref{eqn:xmcmc gamma bayes}), but by using (\ref{eqn:p(y|x*) estimate}) and (\ref{eqn:weighted gaussian}), an approximation of (\ref{eqn:dk* weighted derive}) can be made
\begin{equation} \label{eqn:dk* weighted estimate}
\begin{split}
\hat{d}^\ko_\text{weighted}
&= \frac{d^\kp(d^\kp)^{-N} + d^\km(d^\km)^{-N}} {(d^\kp)^{-N} + (d^\km)^{-N}} \\
&= \frac{d^\kp+d^\km(d^\kp/d^\km)^{N}}{1+(d^\kp/d^\km)^{N}}
.
\end{split}
\end{equation}

For comparison, it is useful to include the original MCMC method in Section~\ref{sec:mcmc} in this framework. It implicitly uses the approximation
\begin{equation} \label{eqn:dp original}
\hat{d}^\ko_\text{original}=2N\sigma_n^2
.
\end{equation}

\begin{figure}[t]
    \centering
    \vspace{-2mm}
    \includegraphics[trim=0mm 1mm 0mm 0mm,clip,width=3.4in]{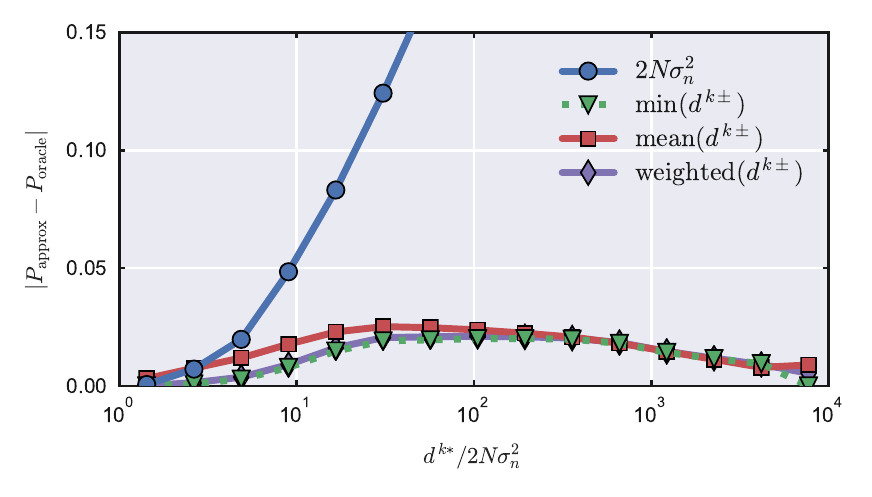}
    \vspace{-1.0\baselineskip}
    \caption{Mean error from using various $d^\ko$ approximations to calculate $P_\text{gibbs}$ with (\ref{eqn:prob gibbs}), (\ref{eqn:gamma sigma_k*}), and (\ref{eqn:sigma_k* oracle}). Parameters: 4 antennas, 64 QAM, $N_\text{gibbs} \!\times\! N_\text{iter} \,\text{=}\, 30\!\times\!30$, WiFi TGn Model-D channel, $E_b/N_0=19\text{dB}$.}
    \label{fig:prob gibbs error}
    \vspace{-.5\baselineskip}  %
\end{figure}

The results of using the {\em min} (\ref{eqn:dp min}), {\em mean} (\ref{eqn:dp mean}), {\em weighted} (\ref{eqn:dk* weighted estimate}), and {\em original} (\ref{eqn:dp original}) estimates of $d^\ko$ to generate $P_\text{gibbs}$ are shown in Fig.~\ref{fig:prob gibbs error}, where the error is the mean of $|P_\text{approx}-P_\text{oracle}|$ over a small range of distances. Based on these plots, the new approximations are all a vast improvement over the {\em original} method. At large distances, all of the new approximations have similar performance since a single bit change has little impact on the total distance, and therefore, $d^\kp \approx d^\km$ resulting in all of the approximations being roughly equivalent. At small distances, the {\em mean} method has the worst performance since it overweights the larger of $d^\kpm$, making the sampler too random.

Since the {\em min} and {\em weighted} approximations have similar performance, the less computationally complex {\em min} option is selected for $\sigma_\ko^2$ in calculating
\begin{equation}\label{eqn:xmcmc gamma sigma_min}
\gamma_k = \frac{d^\km - d^\kp}{\hat{d}_{\min}^\ko/N} + \lambda^{a}_k
.
\end{equation}
The choice of $\hat{d}_{\min}^\ko$ will be verified in Section~\ref{sec:dko approximations} with BER plots in a complete system, but first, output LLR conditioning and pseudo-convergence enhancements will be introduced so that the combined effects can be accounted for.

\subsection{Output LLR Overconfidence and Conditioning}\label{sec:xmcmc - llr output conditioning}

As in the original MCMC algorithm in Section~\ref{sec:mcmc}, the output LLR can be calculated with (\ref{eqn:llr output mcmc}). It uses the list $\text{Z}\!\subset\!\mathbb{X}$ of sampled bit permutations as an approximation to the full permutation list used by the MAP detector in (\ref{eqn:llr output maxlog map}). This is the method used by list based MIMO detectors such as MCMC \cite{Farhang2006}, list sphere-decoding \cite{Hochwald2003}, and K-Best \cite{Guo2006}. Notice that in order to use this method, one must have a list which accurately represents the statistics of the full permutation list. If the list is too short or poorly selected the statistics break down, the calculation becomes unreliable, and the output LLRs can be wildly inaccurate.

One can state that a requirement of the algorithm is for a sufficient number of samples to be taken, but there are two reasons why this is undesirable in practice. First, an implementable design has the goal of minimizing computational complexity and therefore real-world implementations need to use the minimum number of samples possible. As a result, a minority of realizations are likely to be poorly converged, leading to invalid output statistics. Second, many modern communication systems have multiple channel realizations per codeword, for example OFDM. When there are multiple channel realizations, some will be more ill-conditioned than others with a longer convergence time and therefore requiring more Gibbs iterations than the average. During our analysis we have observed situations where the worst channel, often during a deep fade, requires 10x more iterations than average while using a WiFi TGn Model-D channel model, 4 antennas, 64 QAM. Using 10x more iterations consistently is undesirable, but if the slowly converging realizations are halted early while still statistically unstable, then their large incorrect soft-output values can easily corrupt the entire codeword despite comprising a small number of bit errors.

In the context of suboptimal iterative detection and decoding schemes, a constant positive scaling coefficient less than one is sometimes used to scale the extrinsic LLRs \cite{douillard1995iterative,elkhazin2006reduced,colavolpe2011siso}. This has the effect of removing divergence and thus increase stability. Therefore, an alternative to using a large list to deal with outlier difficult channel realizations is to decrease the confidence of their respective LLRs so that they no longer have the strength to corrupt the entire codeword. Unlike other scaling methods which suggest a heuristic constant scaling coefficient to reduce LLR confidence, we propose using the statistics of the sample list to compute a dynamic coefficient that may be used with any list based detector.

If we allow that the sample list may be of poor quality, then there are two contributions to error: the AWGN noise in the system model and the quality of the list. The quality of the list can be computed using the probability that the list contains a sample that is the true transmitted signal without error. As in Section~\ref{sec:xmcmc - gibbs excitation}, if we assume that the combination of both forms of error are Gaussian distributions then the output LLR can be changed to
\begin{equation}\label{eqn:llr output xmcmc}
\begin{split}
\lambda_{k,\text{X-MCMC}}^e \approx 
 &\frac{1}{2}\min_{\mathbf{x}\in\text{Z}^\km} \! \left( \frac{1}{\sigma_z^2}\norm{\mathbf{y}-\mathbf{H}\mathbf{s}^{k^-}}^2 - \mathbf{x}^\wok \!\cdot\! \boldsymbol{\lambda}^{a,\wok} \right) \\
-&\frac{1}{2}\min_{\mathbf{x}\in\text{Z}^\kp} \! \left( \frac{1}{\sigma_z^2}\norm{\mathbf{y}-\mathbf{H}\mathbf{s}^{k^+}}^2 - \mathbf{x}^\wok \!\cdot\! \boldsymbol{\lambda}^{a,\wok} \right)
\end{split}
\end{equation}
where $\sigma_z^2$ represents the variance of the combined contributions of both AWGN noise and list error. This can be estimated with the minimum distance sample in the list similarly to (\ref{eqn:sigma_k* oracle}) using $\hat{d}_\text{min}^\ko$. That is,
\begin{equation} \label{eqn:sigma_z}
\sigma_z^2 \approx \frac{1}{2N} \min_{\mathbf{x}\in\text{Z}} \left(\norm{\mathbf{y}-\mathbf{H}\mathbf{s}}^2\right)
.
\end{equation}
Note that additionally we should limit $\sigma_z^2\ge\sigma_n^2$. This protects against a sample over-fitting the noise and therefore having a distance less than the expectation of correct bit sequences, $2N\sigma_n^2$, which would incorrectly create overconfidence in the output LLR.

The corruption of codewords by rare poorly converged realizations can cause an error floor to appear on BER plots. Using output LLR conditioning reduces this effect and allows a short, fixed number of iterations to be used stably and reliably in real-world applications.

We also note that since the output conditioning is derived using the sample list, this method may also be applicable to other list based algorithms such as list sphere-decoding and K-Best.

\subsection{Summary}\label{sec:xmcmc - summary}

In practice, the X-MCMC algorithm is executed similarly to the original MCMC algorithm outlined in Algorithm~\ref{alg:gibbs}. The changes include the use of the newly introduced dynamic scaling of $(d^\km-d^\kp)$ and output conditioning applied to $\lambda_{k,\text{X-MCMC}}^e$. The full X-MCMC algorithm is outlined in Algorithm~\ref{alg:x-gibbs} with the necessary equations specified. Note that many Gibbs samplers may be used in parallel to increase sampling speed and increase sample diversity. When using parallel Gibbs samplers each $\hat{d}_\text{min}^\ko$ should be calculated independently for each Gibbs sampler whereas the output LLR should be calculated with a list $\text{Z}$ comprising a combination of all samplers. Moreover, $\sigma_z^2$ in (\ref{eqn:sigma_z}) should be obtained by taking the minimum of $\hat{d}_{\min}^\ko$ of all the Gibbs samplers.

\begin{algorithm}
\caption{X-MCMC Gibbs Sampler}
\label{alg:x-gibbs}
\begin{algorithmic}[1]
\State Initialize $\mathbf{x}$.
\For{$1$ to $N_\text{iter}$}
    \For{$k=1$ to $N\log_2(N_\text{qam})$}
        \State Calculate $d^\kp$ and $d^\km$ with (\ref{eqn:dkpm}).
        \State Update $\min_{\mathbf{x}\in\text{Z}}(\;)$ for (\ref{eqn:sigma_z}).
        \State Update $\min_{\mathbf{x}\in\text{Z}^\kp}(\;)$ and $\min_{\mathbf{x}\in\text{Z}^\km}(\;)$ for (\ref{eqn:llr output xmcmc}).
        \State Calculate $P_\text{gibbs}$ with (\ref{eqn:prob gibbs}) and (\ref{eqn:xmcmc gamma sigma_min}).
            \State Generate a uniform random variable $0\le r \le 1$.
            \If{$r < P_\text{gibbs}$}
                \State $x_k \leftarrow \plusone$
            \Else
                \State $x_k \leftarrow \minusone$
            \EndIf
    \EndFor
\EndFor
\State Compute output LLR with (\ref{eqn:llr output xmcmc}), (\ref{eqn:sigma_z}).
\end{algorithmic}
\end{algorithm}

\section{Pseudo-Convergence}\label{sec:pseudo-convergence}

Now that the high SNR stalling problem is solved by the X-MCMC detector and the Gibbs sampler moves efficiently at all SNR levels, a new issue is encountered. The Gibbs sampler may stop moving due to an effect referred to as pseudo-convergence. This problem appears to be less understood in MIMO communications applications, but has been noted in the wider MCMC statistics literature \cite{brooks2011handbook}. Although symptomatically similar, it is different from the stalling problem discussed previously. It occurs when the posterior probability distribution is multi-modal with weak connections between modes, i.e.\ important regions are weakly connected to other important regions through very low probabilities. Thus, the Gibbs sampler may stay in one mode, i.e.\ similar bit permutations connected with sufficient probability, and collect highly correlated samples. When encountered, pseudo-convergence decreases the sampling efficiency of MCMC, resulting in a need for a large number of iterations and/or parallel Gibbs samplers. Others appear to be encountering this effect as reports of performance improvements when adding random walk restarting have been observed \cite{kumar2011near,datta2012novel}.
\subsection{Gibbs Detail Plots}
\begin{figure*}[t]
\begin{minipage}{.48\linewidth}
    \centering
    \vspace{-2mm}
    \includegraphics[trim=0mm 1mm 0mm 0mm,clip,width=\linewidth]{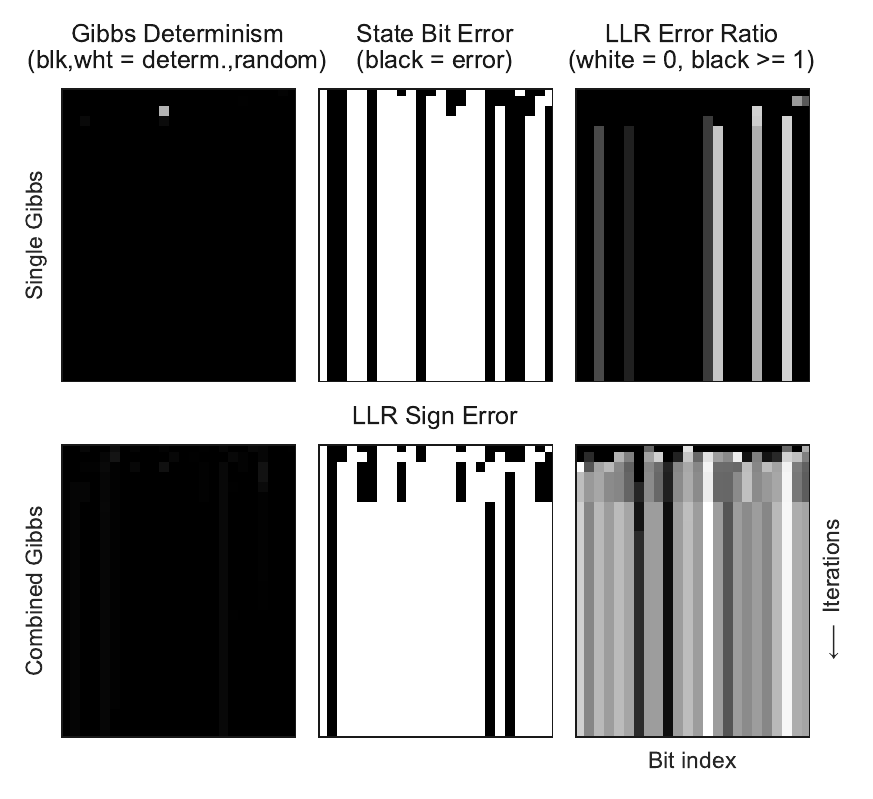}
    \vspace{-1.5\baselineskip}
    \caption{Detail on Gibbs sampler behavior with the original random initialized MCMC. Notice the high SNR stalling problem resulting in no improvement after the first few iterations. Parameters: 4 antennas, 64 QAM, $N_\text{gibbs} \times N_\text{iter} \,\text{=}\, 30\times30$, WiFi TGn Model-D channel, $E_b/N_0=19\text{dB}$.}
    \label{fig:gibbs detail ---}
\end{minipage}
\hfill
\begin{minipage}{.48\linewidth}
    \centering
    \vspace{-2mm}
    \includegraphics[trim=0mm 1mm 0mm 0mm,clip,width=\linewidth]{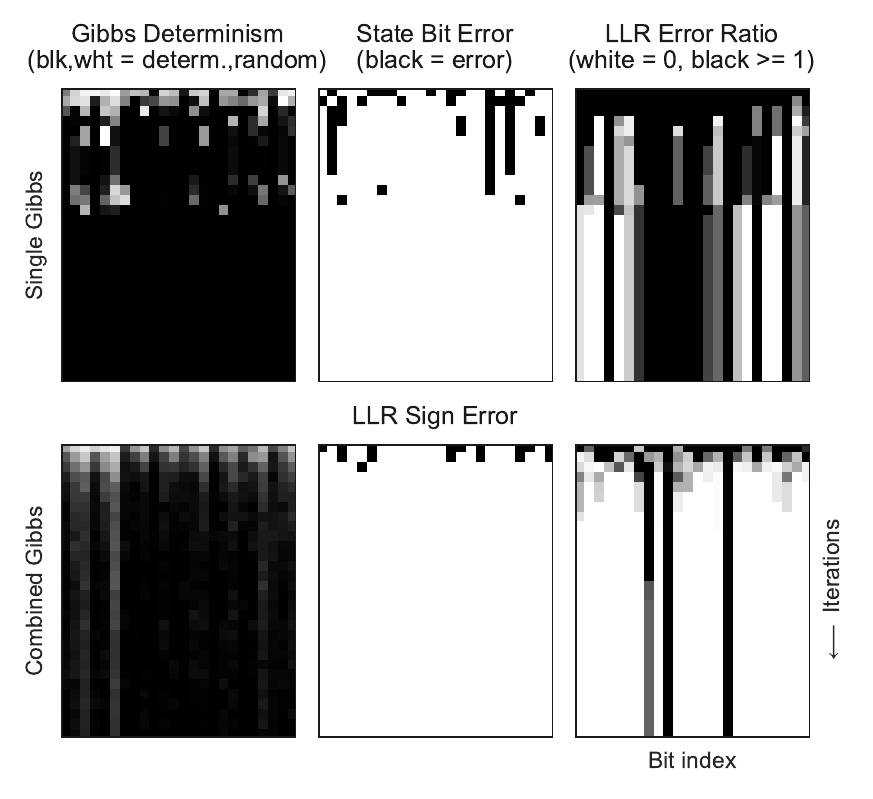}
    \vspace{-1.5\baselineskip}
    \caption{Detail on Gibbs sampler behavior with partial X-MCMC (\mbox{`x - -'} = Gibbs excitement only). Stalling fixed but now there is a pseudo-convergence stopping issue. Parameters: 4 antennas, 64 QAM, $N_\text{gibbs} \!\times\! N_\text{iter} \,\text{=}\, 30\!\times\!30$, WiFi TGn Model-D channel, $E_b/N_0=19\text{dB}$.}
    \label{fig:gibbs detail x--}

\end{minipage}
\vspace{-.5\baselineskip}
\end{figure*}

To understand the pseudo-convergence phenomena more thoroughly, we have developed the Gibbs detail plots of Fig.~\ref{fig:gibbs detail ---}, \ref{fig:gibbs detail x--}, and \ref{fig:gibbs detail x-p}, introduced partially in Section \ref{sec:mcmc - stalling}. For consistency, these figures all use the same input data and initialized states. The first row of subplots show information on a single Gibbs sampler whereas the second row shows information on the combination of all parallel Gibbs samplers of the algorithm run on a single realization of transmitted data.

\begin{equation}\label{eqn:llr error ratio}
\text{LLR error ratio} = |\lambda_k^e-\lambda_{k,\mathrm{MAP}}^e|/\underset{k}{\mathrm{mean}} \left(|\lambda_{k,\mathrm{MAP}}^e|\right)
\end{equation}
 is calculated against the optimal max-log MAP solution. The max-log MAP solution is what MCMC should converge to if run for an infinite number of iterations. In Fig.~\ref{fig:gibbs detail ---}, \ref{fig:gibbs detail x--}, and \ref{fig:gibbs detail x-p}, the LLR error ratio values are shown with a gray-scale color mapping, zero to one as white to black where values above one saturate to black.

The second row shows a combined view of all of the $N_\mathrm{gibbs}$ parallel Gibbs samplers. The first subplot is the average ``determinism'', whereas the second and third are the LLR sign error relative to Max-MAP and the LLR error ratio in (\ref{eqn:llr error ratio}).

The desired behavior of the first column is to show signs of the random walk being guided with a variation of determinism, not fully random or deterministic. In the second column, the single Gibbs sampler should not stay converged to any state and instead should continuously explore the state space, whereas the combined Gibbs samplers should converge to no LLR sign error relative to Max-MAP. Finally, the third column should continue to converge to the Max-MAP solution, displayed by white.

Fig.~\ref{fig:gibbs detail ---} shows that the original MCMC detector described in Section~\ref{sec:mcmc} is almost completely deterministic, is strongly stalled at this high SNR, and does not improve the output LLR after only a few iterations.

By using the excited Gibbs sampler described in Section \ref{sec:xmcmc - gibbs excitation}, Fig.~\ref{fig:gibbs detail x--} shows a large improvement in behavior. It is no longer stalled and after a few Gibbs iterations the algorithm has mostly converged. In the second row of subplots, after an initial very active period, the guided random walk slows as all parallel samplers become locked into an isolated posterior mode due to pseudo-convergence.

\subsection{Detection and Escape}

As pseudo-convergence is a byproduct of the structure of the posterior distribution, the algorithm is behaving correctly and as expected. Given enough iterations a single Gibbs sampler will eventually leave an isolated mode and sample others. Instead of waiting for the low probability of transition, we prefer a more efficient method of detecting when pseudo-convergence has occurred and then forcing state divergence. This allows the algorithm to collect more unique samples with fewer iterations, thus improving the sampling efficiency of the MCMC detector.

Two effective and computationally efficient methods to detect pseudo-convergence include what we refer to as the {\em distance} and {\em motion} methods. The {\em distance} method tracks the best (smallest) distance $d$ sampled over time, including both $d^\kp$ and $d^\km$. If this distance does not improve in $N_\text{motion}$ steps, then pseudo-convergence is detected. Alternatively, the {\em motion} method detects when no change has occurred in Gibbs state $\mathbf{x}$ for $N_\text{motion}$ steps.

The choice of using the $\hat{d}_\text{min}^\ko$ estimate in Section~\ref{sec:xmcmc - gibbs excitation} causes the Gibbs sampler to move slightly more slowly and deterministically, thus we have found that it tends to stop moving when in pseudo-convergence, therefore using the {\em motion} pseudo-convergence detection strategy works well with the choice of $\hat{d}_\text{min}^\ko$. For the detection threshold we use $N_\text{motion}=N \log_2(N_\text{qam})$ steps which is one full Gibbs iteration.

Once pseudo-convergence is detected, the most robust though not necessarily the best solution is to restart the Gibbs sampler with a new fully random state. This is a solution mentioned in the wider MCMC literature beyond MIMO communications applications \cite{brooks2011handbook}. A drawback of the full restart approach is that it requires re-initialization of the Gibbs sampler which may be an expensive and time consuming operation in VLSI implementations. We have found that a full restart is not necessary in a bitwise MCMC detector and under some circumstances is a bad choice. Instead, forcing a 1-bit state change in the next bit following pseudo-convergence detection can be a good solution. This can be thought of as adding an impulse of energy or excitation to the random walk which assists the Gibbs sampler in escaping the isolated posterior mode. Using this 1-bit strategy incurs no additional complexity from re-initializing the sampler and is trivial to implement in VLSI designs.

\begin{figure}[t]
    \centering
    \vspace{-2mm}
    \includegraphics[trim=0mm 1mm 0mm 0mm,clip,width=3.4in]{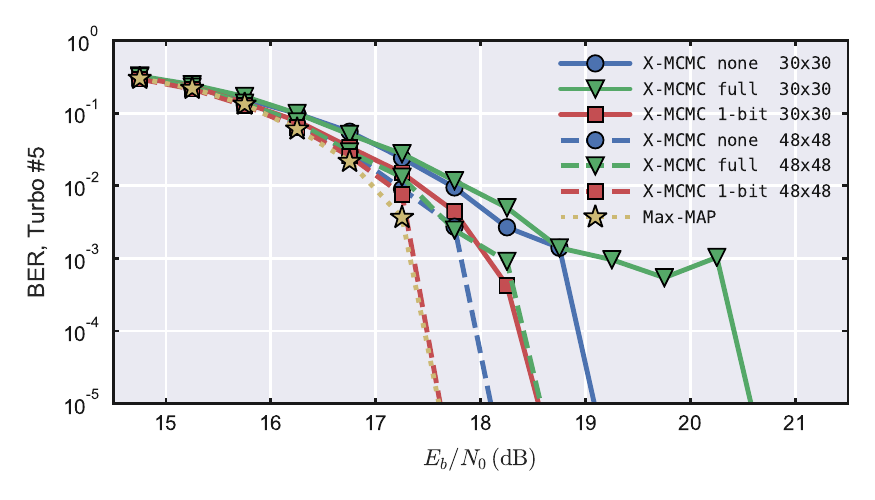}
    \vspace{-1.0\baselineskip}
    \caption{Comparison of using no pseudo-convergence mitigation, full-restart escape, and next-bit forced change escape. Parameters: 4 antennas, 64 QAM.}
    \label{fig:ber pseudoconvergence restart}
    \vspace{-.5\baselineskip}  %
\end{figure}

In Fig.~\ref{fig:ber pseudoconvergence restart} which uses {\em motion} based detection, we see that the full-restart escape method can actually degrade performance. This is due to the number of iterations needed to converge to a region of important samples being larger than the number needed to naturally leave a pseudo-converged region, thus, the full-restart is lowering overall sampling efficiency. For these test parameters, the full-restart creates too large of a divergence which wastes time in re-converging, whereas the next-bit forced change is sufficient to leave the pseudo-converged region without incurring a re-convergence penalty.

\begin{figure}[t]
    \centering
    \vspace{-2mm}
    \includegraphics[trim=0mm 1mm 0mm 0mm,clip,width=3.4in]{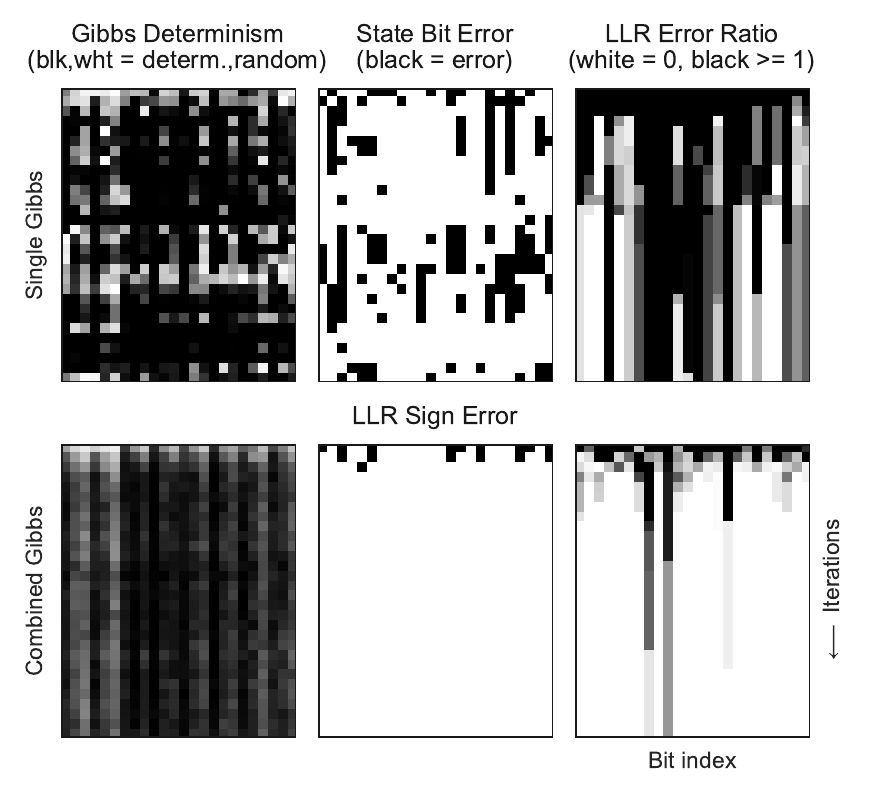}
    \vspace{-1\baselineskip}
    \caption{Detail on Gibbs sampler behavior with partial X-MCMC (\mbox{`x - p'} = Gibbs excitement and pseudo-convergence stopping mitigation). Stalling and stopping are now resolved. Parameters: 4 antennas, 64 QAM, $N_\text{gibbs} \!\times\! N_\text{iter} \,\text{=}\, 30\!\times\!30$, WiFi TGn Model-D channel, $E_b/N_0=19\text{dB}$.}
    \label{fig:gibbs detail x-p}
    \vspace{-.5\baselineskip}
    \vspace{-.5\baselineskip}  %
\end{figure}

By adding the {\em motion} based pseudo-convergence detection and next-bit forced change method to the excited Gibbs sampler we see the results presented in Fig.~\ref{fig:gibbs detail x-p}. Now the stalling problem seen in Fig.~\ref{fig:gibbs detail ---} and the stopping problem seen in Fig.~\ref{fig:gibbs detail x--} are resolved. Both the single and combined Gibbs determinism subplots show that the MCMC algorithm is consistently excited. The combined LLR quickly converges to a correct output bit sequence and then continues to improve the output LLR until near Max-MAP performance is achieved.

\section{Results}\label{sec:results}

One of the most important features of the results that follow is that the WiFi TGn Model-D channel \cite{ieee2004tgn} has been used for $\mathbf{H}$. This realistic, correlated indoor model creates a much more challenging problem compared to the typical Gaussian i.i.d. channels commonly used in the literature \cite{Hochwald2003,Guo2006,Farhang2006,Laraway2009,hassibi2014optimized}. The main issue with using an unrealistically uncorrelated, and therefore unrealistically easy model, is that convergence is relatively easy to achieve, making stalling and stopping issues appear less severe. For more details on channel model selection, see the analysis and testbed results in \cite{hedstrom2017icc}.

We have found that producing BER curves using only the hard-decision of the MIMO detector by itself is insufficient to analyze MCMC performance. Since the MCMC detector can potentially have statistical stability issues, assessing the quality of the soft-output information is essential, therefore, we have used the LDPC 3/4 rate 1944 block length code from the 802.11n specification in generation of all of the following BER curves. It was selected as a moderate coding level among the 1/2, 2/3, 3/4, and 5/6 rates available in the WiFi specification.

For most of the BER results, five turbo loops are used, see Fig.~\ref{fig:turbo_loop}, which enhances final BER performance by allowing the detector and decoder to iteratively exchange extrinsic soft-information \cite{robertson1997optimal}. Although not all applications have sufficient time in their latency budget to do turbo iterations, it is useful to do most of the analysis using these iterations because it allows for the testing of an algorithm's ability to use prior information correctly.

The figures in this section include an MMSE initialized MCMC detector for reference as described in Section~\ref{sec:mcmc - stalling}. The randomly initialized version is not generally included as it performs worse than the MMSE initialized version under all conditions. Max-MAP/Max-ML is shown as the optimal performance bound when possible since the channel capacity for non-Gaussian models is generally unknown. By using a highly optimized GPU implementation we are able to compute MAP/ML at up to 4 antennas with 64 QAM. For the 8 antenna with 256 QAM case we use a very large K-Best as an approximation of the Max-MAP limit since it is known to have near-MAP performance \cite{Guo2006}. One moderate sized K-Best is generally included so that the reader may do some initial comparisons with the literature on K-Best.
\subsection{$d^\ko$ Approximations}\label{sec:dko approximations}

\begin{figure}[t]
    \centering
    \vspace{-2mm}
    \includegraphics[trim=0mm 1mm 0mm 0mm,clip,width=3.4in]{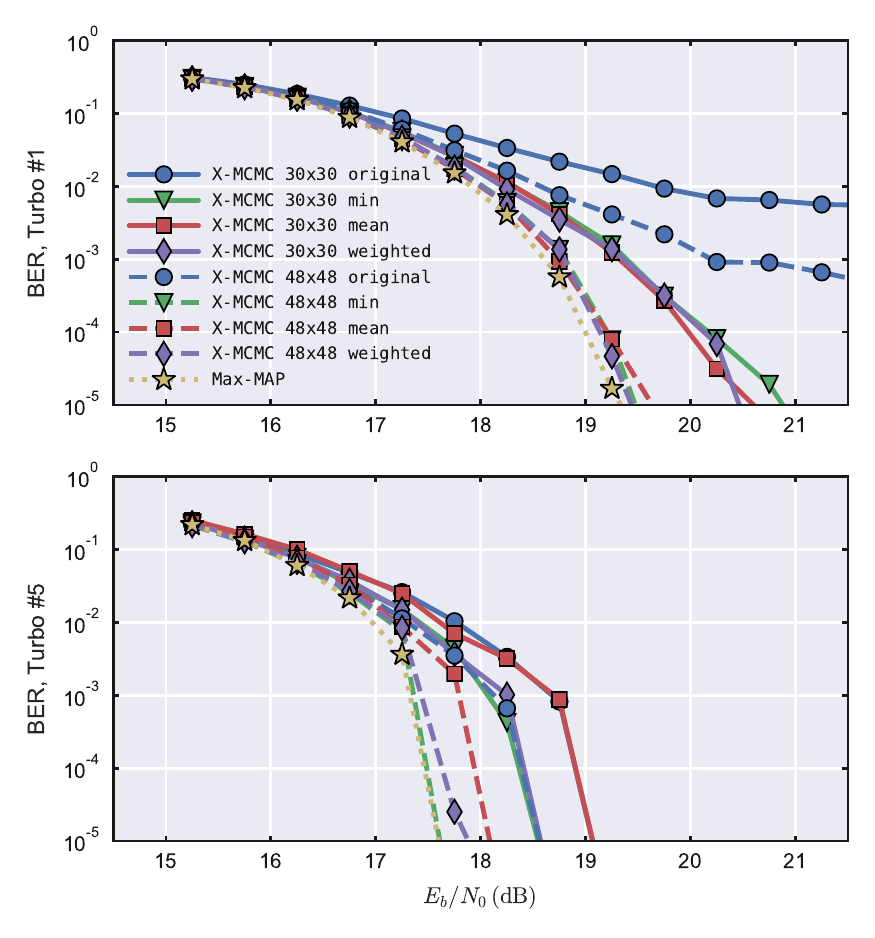}
    \vspace{-1.0\baselineskip}
    \caption{Comparison of $d^\ko$ approximations (\ref{eqn:dp original}), (\ref{eqn:dp min}), (\ref{eqn:dp mean}), and (\ref{eqn:dk* weighted estimate})  with all X-MCMC components enabled including Gibbs excitation, output LLR conditioning, and pseudo-convergence mitigation. Two plots are shown, the top without turbo iterations (Turbo\#1) and with five turbo iterations (Turbo\#5). Both are important as they show the effects without and with use of {\em a priori} information. Parameters: 4 antennas, 64 QAM.}%
    \label{fig:ber sig2e comparison}
    \vspace{-.5\baselineskip}  %
\end{figure}

In the previous sections, the X-MCMC algorithm was developed with greatly enhanced performance. Now with all three components fully developed, we may revisit the selection of $\hat{d}_\text{min}^\ko$ made in Section~\ref{sec:xmcmc - gibbs excitation}. The reason for doing this verification after the development of the other X-MCMC enhancements is that their combined interaction can impact the final choice. Therefore, in Fig.~\ref{fig:ber sig2e comparison}, we show BER plots comparing the suggested approximations to $d^\ko$ while also using the output LLR conditioning and pseudo-convergence enhancements. Both the first and final turbo iterations need to be shown to check for issues with the $\sigma_\ko^2$ scaling relative to the prior $\boldsymbol{\lambda}^a$ in (\ref{eqn:xmcmc gamma sigma_min}).

The BER curves show that the new approximations perform well and will approach near MAP performance given sufficient iterations, though not with equal speed or efficiency. As was also shown in the $P_gibbs$ error analysis of Fig.~\ref{fig:prob gibbs error}, we find that the {\em min} approximation performs well. It remains the preferred approximation since it also has a much lower computational complexity compared to {\em weighted}.

\subsection{EXIT Chart}\label{sec:results - exit}

\begin{figure}[t]
    \centering
    \vspace{-.7mm}  %
    \includegraphics[trim=0mm 1mm 0mm 0mm,clip,width=3.4in]{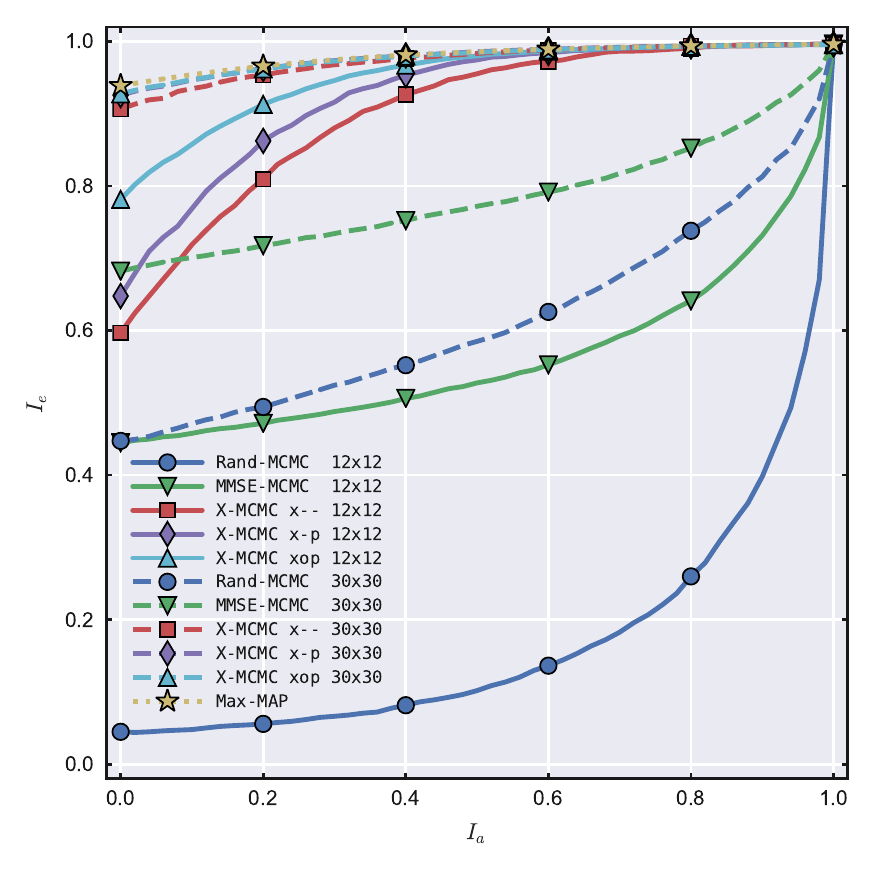}
    \vspace{-1.0\baselineskip}
    \caption{EXIT chart showing random and MMSE initialized original MCMC methods versus X-MCMC components (\mbox{`x o p'} are flags representing inclusion of Gibbs excitation, output LLR conditioning, and pseudo-convergence mitigation). Parameters: 4 antennas, 64 QAM, $E_b/N_0=19\text{dB}$, no coding.}
    \label{fig:exit xop comparison}
    \vspace{-.5\baselineskip}  %
\end{figure}

To deepen the understanding of the MCMC and X-MCMC detectors, an extrinsic information transfer (EXIT) chart is presented in Fig.~\ref{fig:exit xop comparison} with prior information randomly generated as described in \cite{ten2001convergence}. The EXIT chart is useful in evaluating MCMC performance since it is independent of code choice and shows both the ability of the detector to use input soft-information and generate output soft-information \cite{chen2010bitwise,senst2010performance}. Given an amount of input information $I_a$ it shows how much extrinsic output information $I_e$ a given method is able to produce, therefore, EXIT charts provide complementary analysis to detector-only BER curves which do not show the quality of LLR output.

As expected, the output extrinsic information $I_e$ of the X-MCMC detector improves as the excited Gibbs sampler, output LLR conditioning, and pseudo-convergence enhancements are included in the algorithm. The introduction of the excited Gibbs sampler is the most important contribution as it fixes the unusual EXIT curve shapes presented by the random and MMSE initialized original MCMC methods, caused by high SNR stalling. This unusual shape has also been observed in \cite{senst2011rao}. The prolonged, flat shape with a sharp rise at the end of the curves is produced by the large $d^\ko$ underestimate from using (\ref{eqn:dp original}). This means that the $(d^\km-d^\kp)/(\hat{d}_\text{original}^\ko/2N)$ is far overweighted compared to the prior $\boldsymbol{\lambda}^a$ in (\ref{eqn:gamma sigma_k*}). Once $I_a>0.8$ the prior becomes strong enough to overcome the imbalance at this SNR. Because an MMSE initiliazation only provides the algorithm with a good starting point, it does not change the underlying problem.

\subsection{Convergence}\label{sec:results - convergence}

\begin{figure}[t]
    \centering
    \vspace{-2mm}
    \includegraphics[trim=0mm 1mm 0mm 0mm,clip,width=3.4in]{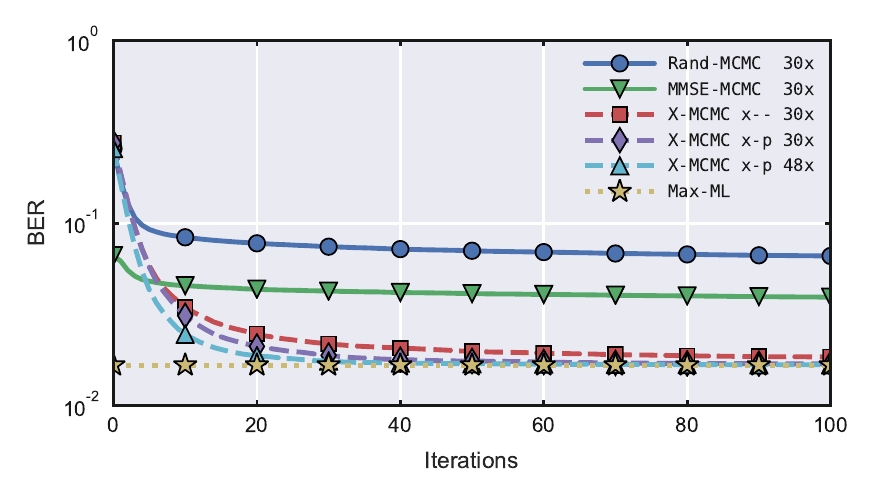}
    \vspace{-1.0\baselineskip}
    \caption{Convergence analysis showing that X-MCMC converges quickly to the ML solution. The BER curves shown use the hard-decision output of the MIMO detector without coding. The number of parallel Gibbs samplers is specified in the legend as `\#x'. The ML/MAP method has been placed as a reference line, though it has no iterations. For X-MCMC, \mbox{`x - p'} are flags representing inclusion of Gibbs excitation and pseudo-convergence mitigation. Parameters:  4 antennas, 64 QAM, $E_b/N_0=19\text{dB}$, no coding.}
    \label{fig:convergence}
    \vspace{-.5\baselineskip}  %
\end{figure}

One of X-MCMC's primary advantages is that it has a fast convergence rate because it has no high SNR stalling problem and avoids pseudo-convergence. This can be seen in Fig.~\ref{fig:convergence} where the hard-decision BER of several detectors are compared as the number of Gibbs iterations increases. The convergence time will generally become longer as the number of antennas increase, the constellation size becomes larger, and the channel becomes more correlated. Note that the output LLR conditioning of X-MCMC is not shown in the figure because scaling does not change the LLR sign and therefore has no effect on the detector BER.

\subsection{BER Performance}\label{sec:results - ber}

\begin{figure}[t]
    \centering
    \vspace{-2mm}
    \includegraphics[trim=0mm 1mm 0mm 0mm,clip,width=3.4in]{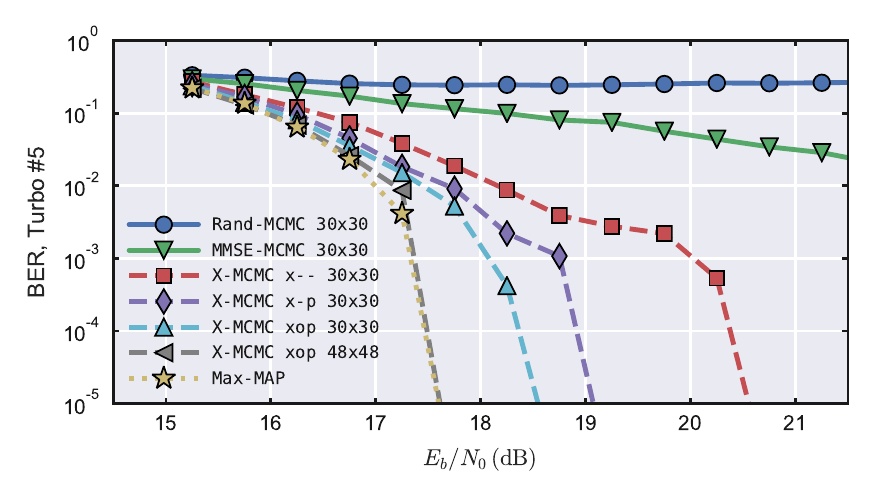}
    \vspace{-1.0\baselineskip}
    \caption{BER curves comparing MCMC methods versus X-MCMC (\mbox{`x o p'} are flags representing inclusion of Gibbs excitation, output LLR conditioning, and pseudo-convergence mitigation). Parameters: 4 antennas, 64 QAM.}
    \label{fig:ber xop comparison}
    \vspace{-.5\baselineskip}  %
\end{figure}

The BER curves of Fig.~\ref{fig:ber xop comparison} confirm the relationships shown in the EXIT chart of Fig.~\ref{fig:exit xop comparison}. There is an incremental improvement in performance as each of the Gibbs excitation, pseudo-convergence enhancement, and output LLR conditioning are included. As predicted by the EXIT chart, a 48$\times$48 X-MCMC detector achieves near Max-MAP performance, and a smaller 30$\times$30 detector is within 1dB. The most interesting feature of these curves is the error floor seen in the X-MCMC curves without output LLR conditioning. This is caused by rare realizations with slow convergence that poorly converge with the fixed number of Gibbs iterations provided. The LLR overconfidence in the poorly converged cases are capable of corrupting entire codewords even when representing a small minority of realizations. For more details on how output LLR conditioning resolves this effect see Subsection~\ref{sec:xmcmc - llr output conditioning}.

\begin{figure}[t]
    \centering
    \vspace{-2mm}
    \includegraphics[trim=0mm 1mm 0mm 0mm,clip,width=3.4in]{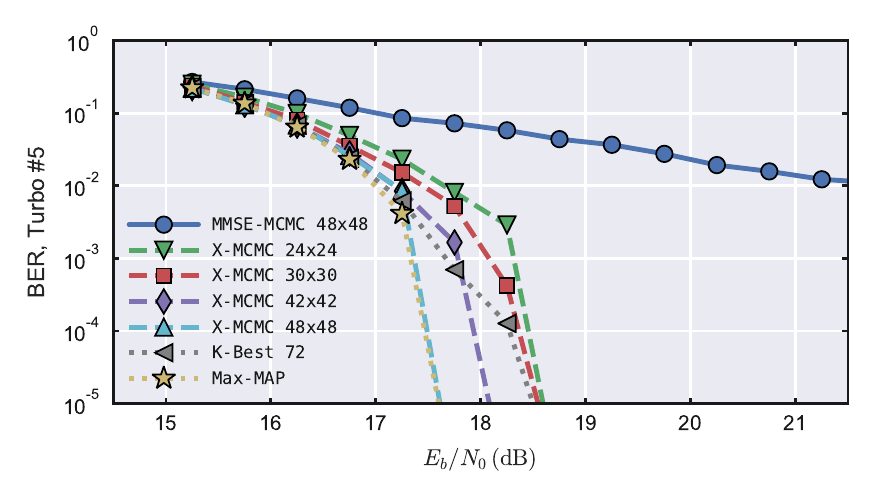}
    \vspace{-1.0\baselineskip}
    \caption{BER curves showing near Max-MAP performance for X-MCMC. Parameters: 4 antennas, 64 QAM.}
    \label{fig:ber 4@64 size}
    \vspace{-.5\baselineskip}  %
\end{figure}

\begin{figure}[t]
    \centering
    \vspace{-2mm}
    \includegraphics[trim=0mm 1mm 0mm 0mm,clip,width=3.4in]{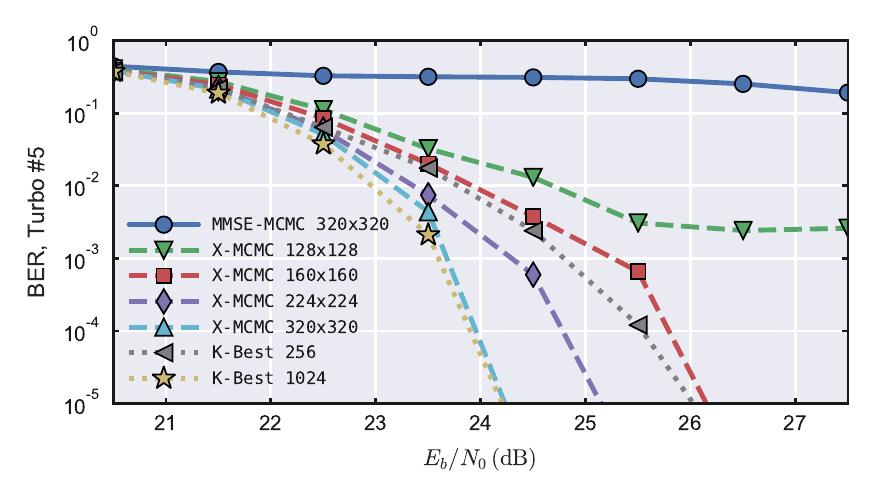}
    \vspace{-1.0\baselineskip}
    \caption{BER curves showing X-MCMC achieves near Max-MAP performance (approximated with large K-Best) even at the maximum 802.11ac MIMO and QAM sizes. Parameters: 8 antennas, 256 QAM.}
    \label{fig:ber 8@256 size}
    \vspace{-.5\baselineskip}  %
\end{figure}

The most important observation in the remaining BER figures is that the X-MCMC detector is capable of achieving near Max-MAP performance under all conditions tested. This is especially impressive at the maximum 802.11ac WiFi protocol size of 8 antenna MIMO with 256 QAM modulation shown in Fig.~\ref{fig:ber 8@256 size}. Compared to MMSE-Initialized MCMC there is a massive $>6$ dB improvement. Similar results are seen in Fig.~\ref{fig:ber 4@64 size} with X-MCMC again achieving Max-MAP performance.

\begin{figure}[t]
    \centering
    \vspace{-2mm}
    \includegraphics[trim=0mm 1mm 0mm 0mm,clip,width=3.4in]{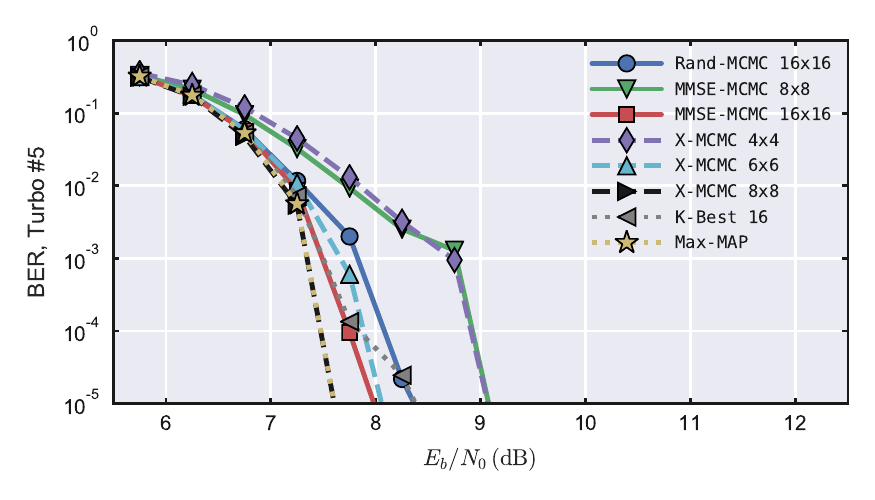}
    \vspace{-1.0\baselineskip}
    \caption{BER curves showing that all methods function at low SNR with small constellation sizes, though X-MCMC is more efficient than previous MCMC methods. Parameters: 4 antennas, 4 QAM.}
    \label{fig:ber 4@4 size}
    \vspace{-.5\baselineskip}  %
\end{figure}

It is relatively easy to achieve near Max-MAP performance on low-order modulation with low-SNR, as seen in Fig.~\ref{fig:ber 4@4 size} and reported in \cite{mao2007markov} and \cite{hedstrom2015markov}. Though MMSE-MCMC works under these conditions, it is at a lower efficiency than X-MCMC. This is predicted by our excited Gibbs derivation since the poor approximation $\hat{d}_\text{original}^\ko$ in (\ref{eqn:dp original}) becomes more accurate at lower SNRs and therefore the impact of stalling is limited.

\subsection{Complexity Analysis}\label{sec:results - complexity}

The proposed X-MCMC algorithm can be broken into four components for complexity analysis: frontend processing (FEP), backend processing (BEP), total number of bitwise steps including all parallel Gibbs samplers, and each bitwise step. To quantify complexity we will use the total number of multiplications as a metric. One can include more complex operations in the analysis, such as exponential functions and division by assuming a roughly equivalent number of multiplies \cite{Guo2006,amiri2007architecture}. For the inverted exponential function used in the probability calculation of (\ref{eqn:prob gibbs}), in \cite{Laraway2009} it is shown that a lookup table (LUT) and adder can provide sufficient accuracy, therefore, its complexity is sufficiently less than a multiplier to be ignored. For the scaling operations of (\ref{eqn:xmcmc gamma sigma_min}) and (\ref{eqn:llr output xmcmc}), we will set the division equal to $C_\mathrm{div}$ multiplies. 

The FEP is used to calculate precomputed values, including $\mathbf{H}^H\mathbf{H}$ and $\mathbf{H}^H\mathbf{y}$, which simplifies each Gibbs sampler step \cite{Laraway2009}. Similarly, the BEP includes the calculation of the output LLR (\ref{eqn:llr output xmcmc}). Thus the FEP and BEP complexity can be represented as $C_\mathrm{FEP} = (2N)^3+(2N)^2$ and $C_\mathrm{BEP} = C_\mathrm{div}+NM$, where $N$ is the number of antennas, and $M$ is the number of bits in each complex constellation symbol. Note that the complex valued matrix $\mathbf{H}$ can be represented by a real version with dimension $2N\times2N$ and likewise for $\mathbf{y}$ with dimension $2N\times1$. The BEP includes a single reciprocal operation to compute the scaling factor which is assumed to be equivalent to a division and then it is applied $NM$ times.

The Gibbs sampler includes the total number of Gibbs sampler bitwise steps multiplied by the complexity of each step as in
\begin{equation}
C_\mathrm{gibbs} = (NM (2NM)^2) \times (2N+1+C_\mathrm{div})
\end{equation}
where we assume that there are $NM$ bitwise steps per iteration, $2NM$ iterations per Gibbs sampler, and $2NM$ parallel Gibbs samplers are used. This number of iterations and parallel samplers is sufficient to achieve near-MAP performance, as shown in Fig.~\ref{fig:ber 4@64 size}, depending on the channel and coding. Using the implementation of \cite{Laraway2009} as a foundation for the X-MCMC enhancements, the step calculation is dominated by a dot product of length $2N$ and the division introduced by our new excitation scaling factor $\sigma_\ko^2$ in (\ref{eqn:xmcmc gamma sigma_min}). It is assumed that the simple state machine necessary to implement the pseudo-convergence strategy of Section~\ref{sec:pseudo-convergence} is of negligible complexity as it requires no arithmetic operations.

\begin{figure}[t]
    \centering
    \vspace{-2mm}
    \includegraphics[trim=0mm 1mm 0mm 0mm,clip,width=3.4in]{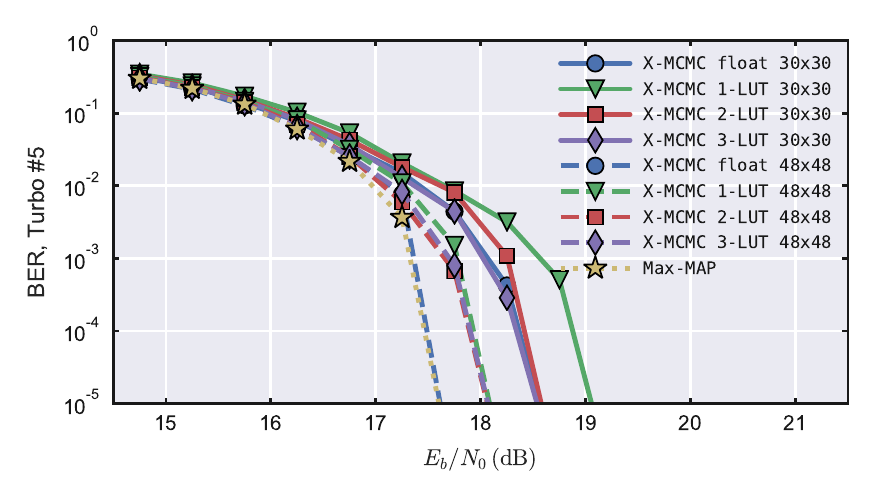}
    \vspace{-1.0\baselineskip}
    \caption{Comparison of using full floating-point precision division vs a simple shift and LUT to replace the scaling operations in (\ref{eqn:xmcmc gamma sigma_min}) and (\ref{eqn:llr output xmcmc}). Note that the 'X-MCMC float 30x30' curve is underneath the corresponding 3-LUT curve. Parameters: 4 antennas, 64 QAM.}
    \label{fig:ber division}
    \vspace{-.5\baselineskip}  %
\end{figure}

It is clear that the division operation introduced to scale the calculation of $\gamma$ in (\ref{eqn:xmcmc gamma sigma_min}) can potentially have a large impact on the leading complexity terms $(8N^4M^3+4N^3M^3C_\mathrm{div}$), therefore, we have explored a simple estimation of the division using a shift unit and n-bit LUT using the most significant bits of the scaling term. In Fig. \ref{fig:ber division}, it is shown that a 3-bit LUT is sufficient to nearly match the floating-point division performance. This is not surprising as the scaling terms themselves are only rough approximations of the local and final error variances.

Since using a rough division approximation is acceptable, we conclude that a division complexity of $C_\mathrm{div}\approx1$ is appropriate. After this simplification, the X-MCMC algorithm adds little additional complexity to each Gibbs step compared to a standard MCMC detector while also significantly reducing the number of needed steps. Thus, the algorithm complexity appears to be polynomial in $N$ and scalable to large MIMO sizes.
\section{Conclusion}\label{sec:conclusion}

We have presented a new derivation of the MCMC detector which solves the high SNR stalling problem without use of hybridization or heuristic temperature scaling terms. Output LLR quality has been improved for poorly converged cases by conditioning output confidence on sample list statistics. Output LLR conditioning is shown to moderate soft-output overconfidence and allow a low complexity fixed length Gibbs sampler to be used in practice, eliminating error floors caused by rare slowly converging realizations. This conditioning may have application to other list based detectors such as list sphere-decoding and K-Best. Additionally, we have identified pseudo-convergence conditions which lower efficiency. The proposed 1-bit randomization procedure is shown as a low complexity alternative way to leave pseudo-convergence compared to using a full random-walk restart. Results show that the combined improvements allow near Max-MAP performance at all SNR regimes with large numbers of antennas and high-order modulation. This is true even with highly correlated, WiFi TGn Model-D channels which are significantly more challenging than the Rayleigh channels with uncorrelated i.i.d.\ Gaussian elements commonly used in the literature. No heuristic optimizations are needed, making the proposed method straightforward to effectively implement in practice. A brief complexity analysis demonstrated that the X-MCMC enhancements requires little additional complexity compared to previous MCMC detectors while dramatically improving performance. A VLSI implementations should be possible with straightforward extensions of existing work.

MIMO sizes beyond our verified 8$\times$8 sizes should be possible within the constraints of the suggested polynomial complexity growth of the X-MCMC algorithm. More correlated channels than the WiFi TGn Model-D channels are possible \cite{hedstrom2017icc}, though with somewhat longer convergence times requiring more Gibbs iterations.

\bibliographystyle{IEEEtran} %
\bibliography{mcmc.bib} %

\end{document}